\documentclass[pra,aps,groupaddress,showpacs]{revtex4}
\usepackage{epsfig,amsmath,graphicx,amssymb,overpic}
\usepackage{bm}
\usepackage{mathrsfs}
\usepackage{graphicx}
\usepackage{epsfig}
\usepackage{dcolumn}
\usepackage{bm}
\usepackage{amsmath,amssymb,amsthm}
\usepackage[colorlinks=true,linkcolor=blue]{hyperref}
\usepackage{subfigure}
\usepackage{booktabs}
\usepackage[mathscr]{euscript}
\usepackage{ulem}
\usepackage{subfigure}
\newcommand{\bea}{\begin{eqnarray}}
\newcommand{\eea}{\end{eqnarray}}
\newcommand{\beq}{\begin{equation}}
\newcommand{\eeq}{\end{equation}}

\def\/{\over}

\usepackage{CJK}
\begin{document}
\begin{CJK*}{GBK}{song}
\title{Special decompositions and linear superpositions of nonlinear systems:\\ BKP and dispersionless BKP equations}
\author{Xiazhi Hao$^1$ and S. Y. Lou$^2$\thanks{Corresponding author:lousenyue@nbu.edu.cn. Data Availability Statement: The data that support the findings of this study are available from the corresponding author
upon reasonable request.}}
\thanks{Corresponding author: lousenyue@nbu.edu.cn. Data Availability Statement: The data that support the findings of this study are available from the corresponding author
	upon reasonable request.}
\affiliation{\footnotesize{$^1$College of Science, Zhejiang University of Technology, Hangzhou, 310014, China}\\
\footnotesize{$^2$School of Physical Science and Technology, Ningbo University, Ningbo, 315211, China}}

\begin{abstract}
The existence of decomposition solutions of the well-known nonlinear BKP hierarchy is explored. 
It is shown that these decompositions provide simple and interesting relationships between classical integrable systems and the BKP hierarchy.
Further, some special decomposition solutions display a rare property: they can be linearly superposed.
With the emphasis on the case of the fifth BKP equation, the structure characteristic having linear superposition solutions is analyzed. Finally, we obtain similar superposed solutions in the dispersionless BKP hierarchy.
\end{abstract}

\maketitle

\section{Introduction}
The linear superposition principle is encountered in many branches of physics and can be employed to solve essentially all linear problems \cite{ks12}. This principle states that, for linear systems, superposition provides large classes of solutions and gives rise to generalized solutions of linear problems in the form of a linear combination of the independent solutions. 
When linear superposition holds, a system can be decomposed into its constituent parts and the behavior of each component is independent of other components \cite{bishop11}.
Conversely, when linear superposition fails, such systems often exhibit behaviors reflecting the fact that individual system components are not independent of each other. 
Linear superposition normally operates in linear systems.
Instead, the lack of linear superposition is one of the crucial characteristics of nonlinear systems.
The loss of linear superposition also has implications for causation, reduction, emergence, and natural laws in nonlinear dynamics, all of which raise important issues for the application of nonlinear models to real-world problems.
Generally, because of the nonlinear terms, linear superposition principle does not hold well in nonlinear theories even for nonlinear integrable systems where the nonlinear superposition is valid.
Nonetheless, it was found to apply for specific cases, such as periodic solutions \cite{Cooper5,pa7,pramana4,mm9,mf8,ag6,km10}. 
The main property that allows for the application of linear superposition principle to these specific nonlinear cases is the reduction of the nonlinear cross terms into linear ones, which then combine with other linear terms \cite{sk11,base,base1}. This is possible only for certain types of solutions with this property.

Here, we will explicitly show a different method allowing for suitable linear combinations of special decomposition solutions which leads to completely different types of solutions.
What is even more remarkable is that the decompositions allow us to uncover unexpected relations between various different integrable systems.
To illustrate our approach, we begin with the B-type Kadomtsev-Petviashvili equation (BKP).



The BKP equation 
\begin{equation}
u_{xt}+(u_{x4}+15uu_{x2}+15u^3-15uv-5u_{xy})_{xx}-5u_{yy}=0,\ v_x=u_y, \label{BKP}
\end{equation}
where $u_{x}=\partial_xu,\ u_{x2}=\partial_x^2u,\ u_{x3}=\partial_x^3u,\ \ldots,\ $
is one of the most mystery models possessing many elegant properties \cite{BKP1,BKP2}. For further calculations, it is convenient to consider the potential form of \eqref{BKP} by setting $u=w_x$,
\begin{equation}
w_{xt}=5w_{yy}-(w_{x5}+15w_xw_{x3}+15w_x^3-15w_xw_y-5w_{xxy})_{x}\equiv K_x. \label{PBKP}
\end{equation}
When $u$ is $y$ independent or $u_y\sim au_x$, the BKP equation \eqref{BKP} is returned to the known Sawada-Kotera (SK) equation \cite{SK, SK1, SK2} which can be widely applied to many physical fields where the KdV equation is valid and the higher-order perturbations need to be considered \cite{JPC}.

In the next section, the potential BKP (PBKP) hierarchy is briefly rewritten down employing the mastersymmetry method (MM) \cite{Master,Master1,ChengYi} which is equivalent to the formal series symmetry approach (FSSA) \cite{FSSA,FSSA1,FSSA3}.
Then, we detail the steps leading to some special types of decompositions of the PBKP hierarchy in section 3. These decompositions
provide simple and interesting relationships for classical integrable systems and are fundamental for subsequent construction of linear superposition solutions.
In section 4, we show that suitable linear superpositions of some special decomposition solutions obtained in section 3 of the PBKP hierarchy are still in fact new solutions of the same equations.  Using these linear superposition formulas, we find several types of exact solutions including $m$+$n$ solitons, $n$ solitons with periodic cnoidal background waves, $n$ soliton solutions combined with soliton-cnoidal wave interaction solutions, and combination solutions of two different modified Schwarzian KdV waves, etc. of the fifth PBKP equation.
In section 5, we present that similar special types of decompositions and linear superpositions still work for the dispersionless PBKP (dPBKP) hierarchy.
Section 6 is devoted to conclusions.

\section{PBKP hierarchy via mastersymmetry method}
A symmetry, $\sigma,$ of the PBKP equation \eqref{PBKP} is defined as a solution of
\begin{equation}\label{K'}
\sigma_{xt}=\partial_x K'\sigma \equiv \partial_x\left(5\partial_x^{-1}\partial_y^2
-\partial_x^5-15w_{x3}\partial_x-15w_x\partial_x^3-45w_x^2\partial_x
+15w_x\partial_y+15w_y\partial_x+5\partial_x^2\partial_y\right)\sigma,
\end{equation}
which means \eqref{PBKP} is invariant under the infinitesimal transformation $w\longrightarrow w+\epsilon \sigma$ with infinitesimal parameter $\epsilon$. If $\sigma$ is not explicitly $t$ dependent, then the symmetry definition equation \eqref{K'} is equivalent to
\begin{equation}\label{KS'}
K_{[,]}\sigma=0,
\end{equation}
where the commutate operator $K_{[,]}$ is defined as
\begin{equation}\label{Kf'}
K_{[,]}f\equiv K'f-f'K\equiv \lim_{\epsilon\rightarrow0}\frac{\rm d}{\rm d\epsilon} \left[K(w+\epsilon f)-f(w+\epsilon K)\right]
\end{equation}
for arbitrary $f$.\\
\bf Conjecture 1. \rm
\begin{equation}\label{Cj1}
K_{2n-1}=\frac{1}{3\cdot 5^n n!}K_{[,]}^{n}y^{n},\quad n=1,\ 2,\ \ldots,\ \infty
\end{equation}
are all symmetries of the PBKP equation \eqref{PBKP}, that means $y^n$ is a master symmetries for all positive integers $n\geq 1$ and $K_{[,]}^{n+1}y^n=0$.

According to the conjecture 1, the PBKP hierarchy can be written as
\begin{equation}\label{PBKPH}
w_t=K_{2n-1}=\frac{1}{3\cdot 5^n n!}K_{[,]}^{n}y^{n},\quad n=1,\ 2,\ \ldots,\ \infty.
\end{equation}
After finishing cumbersome calculations, we have proved the conjecture 1 for $n= 1,\ 2,\ 3,\ 4$ and $5$ with the following explicit forms,
\begin{equation}\label{K1}
w_{t_1}=K_{1}=\frac{1}{15}K_{[,]}y=w_x,
\end{equation}
\begin{equation}\label{K3}
w_{t_3}=K_{3}=\frac{1}{150}K_{[,]}^2y^2=3w_y,
\end{equation}
\begin{equation}\label{K5}
w_{t_5}=K_{5}=\frac{1}{2250}K_{[,]}^3y^3=5\partial_x^{-1}w_{y2}-w_{x5}-15w_xw_{x3}-15w_x^3 +5w_{yx2}+15w_x w_y=K,
\end{equation}
\begin{eqnarray}\label{K7}
w_{t_7}&=&K_{7}=\frac{1}{45000}K_{[,]}^4y^4\\
&=&-w_{x7} -21 w_{x3}^2-21 w_{x2}w_{x4} -21 w_x w_{x5} +21\partial_x^{-1}(w_xw_{yy})+7\partial_x^{-2}w_{y3}+42 w_{x2}w_{xy}\nonumber\\
&&-42w_{x3}w_y +42\partial_x^{-1}(w_{x4}w_y)-63\partial_x^{-1}(w_x^2w_{xy}) -21w_xw_{yx2}-63w_xw_{xx}^2 -126w_x^2w_{x3} +21w_{xy2}\nonumber\\
&&+21 w_x\partial_x^{-1}w_{y2}-63w_x^4+63/2w_y^2,\nonumber
\end{eqnarray}
\begin{eqnarray}\label{K9}
w_{t_9}&=&K_{9}=\frac{1}{1125000}K_{[,]}^5y^5\\
&=&9\partial_x^{-2}\{\partial_x^{-1}w_{y4}-w_{yx8}
+6 w_{x2y3}-270 w_{x} w_{x2}^2 w_{y}+3 w_{x5y2}+36 w_{yx2}^2+9 w_{x}w_{x3y2} +9 w_{x}w_{y3}\nonumber\\
&&+3(10 w_{xy2}-105 w_{x2}w_{xy}-111 w_{x}w_{yx2}-f) w_{x3} -9(39 w_{x}w_{x2}+5 w_{xy}) w_{yx3}-18 w_xw_{yx6}\nonumber\\
&& +27 w_xw_{xy}^2+15 w_{xy2}w_y-3 w_{x7} w_{y}
-9(21 w_xw_{xy}+10 w_{x2} w_y-w_{y2}+7 w_{yx3}) w_{x4}  +36 w_{x2}w_{x2y2}
\nonumber\\
&&-45 w_{x2}w_{yx5}+9 w_{x2}\partial_x^{-1}w_{y3}-18 w_{x6}w_{xy} +27 w_{xy}w_{y2} +9[2(5 w_y-27 w_x^2) w_{xy} +3 w_xw_{y2}] w_{x2}\nonumber\\
&&+3(5 w_y-27 w_x^2-20 w_{x3}) w_{yx4}-45(w_xw_y+w_{yx2}) w_{x5}
-63 w_{x3}^2 w_y\nonumber\\
&&+9 w_{yx2}(\partial_x^{-1}w_{y2}-12 w{x}^3+5 w_xw_y-27 w_{x2}^2)\},\ \nonumber\\
f_x&=&6(15w_xw_{x2} -2w_{xy}-w_{x4})w_y+54w_x^2w_{xy}-\partial_x^{-1}w_{y3}-3(w_xw_{y})_y.\nonumber
\end{eqnarray}
It should be mentioned that the ninth-order PBKP equation \eqref{K9} is also only a (2+1)-dimensional extension of the seventh-order Sawada-Kotera (SK) equation. Both equations \eqref{K7} and \eqref{K9} will be reduced back to the seventh-order SK equation when we take $w_y=aw_x$.

Though the conjecture is difficult to prove for general $n$, its formal extended form can be restrictly proved by using FSSA \cite{FSSA,FSSA1,FSSA3}. \\
\bf Proposition 1.  \rm The PBKP equation \eqref{PBKP} possesses formal series symmetries
\begin{equation}\label{FS1}
\sigma_{2n-1}=\frac{1}{3\cdot 5^n n!}\sum_{k=0}^{\infty}f^{(n-k)}K_{[,]}^{k}y^{n},\quad n=1,\ 2,\ \ldots,\ \infty,
\end{equation}
where $f$ is an arbitrary function of $t$.

The correctness of proposition 1 has been proved for general $t$-independent $K$ by one of the present authors (Lou) in several earlier works \cite{FSSA,FSSA1,FSSA3,FSSA2}. Thus, conjecture 1 is equivalent to assuming that the formal series \eqref{FS1} is truncated up to $n$ and the special function $f=1$.

\section{Decompositions of the PBKP hierarchy}
By using the symmetry constraint method to the Lax pair of the BKP equation, it is known that the BKP hierarchy can be solved by decomposing the BKP hierarchy to the KdV flows \cite{ChengYi}. After finishing some tedious calculations, we find that for every equation of the PBKP (and then BKP) hierarchy, there are three consistent decompositions from the PBKP hierarchy to potential KdV (PKdV) flows. 
More different ways to relate BKP and KdV hierarchies with reductions were described in \cite{Rela1,Rela2,Rela3}. Another point worth bringing up is that in addition to the KdV hierarchy, we have found relationships between other classical systems and the BKP hierarchy. The relationships between these various models are illustrated below.

\subsection{Decompositions of the fifth-order PBKP equation (\ref{PBKP})}
Usually, for a higher dimensional integrable system, one can formally separate some variables by using symmetry constraints on its Lax pair. For PBKP equation \eqref{PBKP}, the Lax pair possesses the form
\begin{eqnarray}
&&\psi_y+\psi_{x3}+3w_x\psi_x=0,\label{Lx}\\
&&\psi_t-9\psi_{x5}-45 w_x \psi_{x3}-45 w_{x2}\psi_{x2} -15(2 w_{x3}+3w_x^2-w_y)\psi_x=0\label{Lt}.
\end{eqnarray}
It is not difficult to verify that $\psi_x$ is a special symmetry of the PBKP equation, i.e., $\sigma=\psi_x$ is a solution of \eqref{K'}. Thus, substituting the symmetry constraint $\psi_x=w_x$, i.e,
\begin{equation}
\psi=w, \label{su}
\end{equation}
into the Lax pair \eqref{Lx} and \eqref{Lt}, we have
\begin{eqnarray}
&&w_y+w_{x3}+3w_x^2=0,\label{PKdV3}\\
&&w_t-9w_{x5}-45(2 w_{x3}+3w_x^2)w_x=0\label{PKdV5}.
\end{eqnarray}
Because equation \eqref{PKdV3} is not explicitly $t$-dependent and \eqref{PKdV5} is not explicitly $y$-dependent, this kind of decomposition is called nonlinearization \cite{VSA} and also known as variable separation approach \cite{VSA1}. Since $w$ is still a function of $y$ and $t$, we call this kind of method as formally variable separation approach (FVSA) \cite{FVSA,FVSA1}.
The separation equations \eqref{PKdV3} and \eqref{PKdV5} are nothing but the KdV and the fifth-order KdV equations, respectively. That means if $w$ is a solution of the KdV and fifth-order KdV decompositions \eqref{PKdV3} and \eqref{PKdV5}, then it is also a solution of the PBKP equation \eqref{PBKP}.

In fact the FVSA can be applied to nonlinear systems irrelative the existence of Lax pairs \cite{FVSA}, since it can be conjectured that any (2+1)-dimensional nonlinear system has a solution decomposed in the form
\begin{eqnarray}
&&w_y=F(w,\ w_x,\ w_{x2},\ \ldots,\ w_{xm}),\label{wy}\\
&&w_t=G(w,\ w_x,\ w_{x2},\ \ldots,\ w_{xn}) \label{wt}
\end{eqnarray}
with the consistent condition 
\begin{eqnarray}
&&w_{yt}-w_{ty}=[F,\ G]=0 \label{wytwty}
\end{eqnarray} preserved.
So it is not necessary to stipulate the exact forms of the functions $F$ and $G$, it is only necessary 
to specify that \eqref{wy} and \eqref{wt} satisfy the PBKP equation \eqref{PBKP} and the consistent condition \eqref{wytwty}. Then a direct calculation shows the following decomposition theorem. \\
\bf Proposition 2. \rm If $w_1,\ w_2,\ w_3,\ w_4,\ w_5$ and $w_6$ are the solutions of the following decomposed systems
\begin{eqnarray}
&&\left\{\begin{array}{l}
\displaystyle{w_{1y}=(\Phi_1+c) w_{1x}+c_1},\
\Phi_1\equiv \partial_x^2+4w_{1x}-2\partial_x^{-1}w_{1x2}, \\
\\
\displaystyle{w_{1t}=(9\Phi_1^2+15c \Phi_1+5(c^2+3c_1))w_{1x}},
\end{array}\right. \label{DC51}\\
&&\left\{\begin{array}{l}
\displaystyle{w_{2y}=\Phi_2w_{2x}+c_1},\ \Phi_2\equiv \partial_x^2+2w_{2x}-\partial_x^{-1}w_{2x2},\\
\\
\displaystyle{w_{2t}=(9\Phi_2^2+15c_1)w_{2x}},
\end{array}\right. \label{DC52}\\
&&\left\{\begin{array}{l}
\displaystyle{w_{3y}=-\frac12( \Phi_3-2c) w_{3x}+c_1,\
	\Phi_3\equiv \partial_x^2+4w_{3x}-2\partial_x^{-1}w_{3x2},} \\
\\
\displaystyle{w_{3t}=\bigg(5c^2+15c_1-\frac94 \Phi_3^2\bigg)w_{3x}},
\end{array}\right. \label{DC53}\\
&&\left\{\begin{array}{l}
\displaystyle{{w_{4y}=w_{4x3}-\frac34\frac{w_{4x2}^2}{W^2}}+\frac32W^4-c_1= \Phi_4w_{4x}+c_1,\ W^2\equiv w_{4x}+c}, \\
\\
\displaystyle{w_{4t}=9\Phi_4^2w_{4x}+15c_1w_{4x},\ \Phi_4\equiv \partial_x^{-1}{W}\partial_x^2W^{-1}\partial_x +W^2+\partial_x^{-1}W^2\partial_x,}
\end{array}\right. \label{DC5JM}\\
&&\left\{\begin{array}{l}
\displaystyle{w_{5y}=\frac14(4\Phi_5^2+6c\Phi_5+3c^2)w_{5x},\ \Phi_5\equiv \partial_x+\frac12 w_5+\frac12 w_{5x}\partial_x^{-1}},\\
\\
\displaystyle{w_{5t}=\frac9{16}(16\Phi_5^4+40c\Phi_5^3+40c^2\Phi_5^2+20c^3\Phi_5+5c^4)w_{5x}},
\end{array}\right. \label{DCSTO}\\
&&\left\{\begin{array}{l}
\displaystyle{w_{6y}=cw_{6x}+c_1},\\
\\
\displaystyle{w_{6t}=-w_{6x5}+5(c-3w_{6x})w_{6x3}+15cw_{6x}^2 -15w_{6x}^3+5(c^2+3c_1)w_{6x},}
\end{array}\right. \label{DCSK}
\end{eqnarray}
then $w_1,\ w_2,\ w_3,\ w_4,\ w_5$ and $w_6$ are all solutions of the PBKP equation \eqref{PBKP}. \\
\bf Proof. \rm Substituting \eqref{wy} and \eqref{wt} with $m>3$ into the PBKP equation \eqref{PBKP} and the decomposition consistent condition \eqref{wytwty}, one can find that there is no possible decomposition \eqref{wy} and \eqref{wt} with $m>3$. Thus, we just take $m=3$ and then $n=2m-1=5$ in the decomposition relations \eqref{wy} and \eqref{wt}. Substituting \eqref{wy} and \eqref{wt} into \eqref{PBKP}, we have
\begin{eqnarray}
w_{x6}(1+5F_{x_3}^2 -5F_{x_3}+G_{x_5})+W(x_0,\ x_1,\ \ldots,\ x_5)=0, \label{x6}
\end{eqnarray}
where $F=F(w,\ w_x,\ w_{x2},\ w_{x3})\equiv F(x_0,\ x_1,\ x_2,\ x_3)$, $G=G(w,\ w_x,\ w_{x2},\ w_{x3},\ w_{x4},\ w_{x5})\equiv G(x_0,\ x_1,\ \ldots,\ x_5)$ and $W=W(x_0,\ x_1,\ \ldots,\ x_5)$ is a complicated expression of $x_0,\ x_1,\ \ldots,\ x_5$. Vanishing coefficient of $w_{x6}$, we have
\begin{eqnarray}
G=(5F_{x_3}-5F_{x_3}^2-1)x_5+G_1(x_0,\ x_1,\ \ldots,\ x_4), \label{GG1}
\end{eqnarray}
where $G_1=G_1(x_0,\ x_1,\ \ldots,\ x_4)$ is a function of $\{x_0,\ x_1,\ \ldots,\ x_4\}$.
By using the relation \eqref{GG1}, \eqref{x6} is changed to
\begin{eqnarray}
w_{x5}\big[G_{1x_4}-5 (F_{x_3}+2)(x_4F_{x_3x_3}+x_3 F_{x_2x_3}+x_2F_{x_1x_3}+x_1F_{x_0x_3}) +10F_{x_2}F_{x_3}+5F_{x_2}\big]+W_1=0, \label{x5}
\end{eqnarray}
where $W_1=W_1(x_0,\ x_1,\ \ldots,\ x_4)$ is $w_{x5}$ independent. Eliminating the coefficient of $w_{x5}$ yields
\begin{eqnarray}
G_{1}=5 (F_{x_3}+2)(\frac12 x_4F_{x_3x_3}+x_3 F_{x_2x_3}+x_2F_{x_1x_3}+x_1F_{x_0x_3})x_4 +5x_4F_{x_2}(2F_{x_3}+1)+G_2, \label{GG2}
\end{eqnarray}
with $G_2\equiv G_2(x_0,\ x_1,\ x_2,\ x_3)$ being a function of $\{x_0,\ x_1,\ x_2,\ x_3\}$.

Similarly, substituting the decomposition \eqref{wy} and \eqref{wt} with $m=3,\ n=5$ and the results \eqref{GG1} and \eqref{GG2} into the consistent condition \eqref{wytwty}, we have
\begin{eqnarray}
5w_{x7}(1-F_{x_3})^2(x_1F_{x_0x_3} +x_2F_{x_1x_3} +x_3F_{x_2x_3} +{x_4}F_{x_3x_3})+\Gamma=0, \label{x7}
\end{eqnarray}
where $\Gamma=\Gamma(x_0,\ \ldots,\ x_6)$ is a $w_{x7}$ independent function of the lower-order differentiations of $w$ with respect to $x$.

Vanishing the coefficient of $w_{x7}w_{x4}$ in \eqref{x7}, we get
\begin{eqnarray}
F=F_1(x_0,\ x_1,\ x_2)x_3+H(x_0,\ x_1,\ x_2). \label{rF}
\end{eqnarray}
Substituting \eqref{rF} into \eqref{x7} and requiring the coefficient of $w_{x7}$ being zero results $F_1(x_0,\ x_1,\ x_2)=a$, a constant. Thus, we have,
\begin{eqnarray}
F=a x_3+H(x_0,\ x_1,\ x_2). \label{rF0}
\end{eqnarray}
Because of the simplification \eqref{rF0}, \eqref{x5} is simplified to
\begin{eqnarray}
&&w_{x4}\big[G_{2x_3}-5(a+2)x_3 H_{x_2x_2}-5(a+2)x_2 H_{x_1x_2}+5x_1(3-a H_{x_0x_2} -3a-2H_{x_0x_2})\nonumber\\
&&-10a H_{x_1}-5 H_{x_2}^2-5 H_{x_1}\big]+W_2=0 \label{x4}
\end{eqnarray}
with $W_2=W_2(x_0,\ x_1,\ x_2)$. Vanishing the coefficient of $w_{x4}$ in \eqref{x4} leads to
\begin{eqnarray}
G_{2}&=&5\big[\frac{a+2}2x_3 H_{x_2x_2}+(a+2)x_2 H_{x_1x_2}-x_1(3-a H_{x_0x_2} -3a-2H_{x_0x_2})+2a H_{x_1}\nonumber\\
&&+ H_{x_2}^2+ H_{x_1}\big]x_3+J=0, \label{G2}
\end{eqnarray}
where $J=J(x_0,\ x_1,\ x_2)$. Up to now, the decomposition relation is simplified to
\begin{eqnarray}
w_y&=&aw_{x3}+H,\label{wy1}\\
w_t&=&(5a^2+5a-1)w_{x5}+5(2a+1)w_{x4}H_{x_2}+5\big[\frac{a+2}2x_3 H_{x_2x_2}+(a+2)x_2 H_{x_1x_2}\nonumber\\
&&-x_1(3-a H_{x_0x_2}-3a-2H_{x_0x_2})+2a H_{x_1}+ H_{x_2}^2+ H_{x_1}\big]x_3+J=0, \label{wt1}
\end{eqnarray}
with one constant $a$ and two undetermined functions $H=H(x_0,\ x_1,\ x_2)$ and $J=J(x_0,\ x_1,\ x_2)$.

Inserting \eqref{wy1} and \eqref{wt1} into the PBKP equation \eqref{PBKP} and the consistent condition \eqref{wytwty}, then, vanishing the coefficients of $w_{xk}$ for $k\geq 3$ leaves the determining equations on $\{a,\ H,\ J\}$,
\begin{eqnarray}
&&aH_{x_2x_2x_2}=0,\ (1-a)[2(a+2)(x_2H_{x_1x_2x_2}+x_1H_{x_0x_2x_2})+(H_{x_2}^2+3H_{x_1})_{x_2}]=0,\label{DS}\\
&&(1-a)[(a+2)(2x_1x_2H_{x_0x_1x_2}+x_1^2H_{x_0x_0x_2}+x_2^2H_{x_1x_1x_2}) +(1-a)(x_2H_{x_1x_1}+x_1H_{x_0x_1})\nonumber\\
&&+[(a+2)x_2-2x_1H_{x_2}]H_{x_0x_2} -2x_2H_{x_2}H_{x_1x_2}-9ax_2]=0,\nonumber\\
&&(1-a)(x_1H_{x_0x_2}+x_2H_{x_1x_2})=0,\ (1-a)H_{x_2x_2}=0,\ (5a^2-2)\partial_{x_2}^5H=0, \nonumber\\
&&(2H_{x_2}-3ax_1\partial_{x_0}-3ax_2\partial_{x_1})H_{x_2x_2}=0,\ (7a^2-4)\partial_{x_2}^4H=0, \nonumber
\end{eqnarray}
\begin{eqnarray}
&&[(5a^2-2)(x_2\partial_{x_1}+x_1\partial_{x_0})\partial_{x_2}+(7a^2-4)\partial_{x_1} -aH_{x_2}\partial_{x_2}+(6-9a)H_{x_2x_2}]H_{x_2x_2x_2}=0,\label{DS1}\\
&&[3(7a^2-4)(x_2\partial_{x_1}+x_1\partial_{x_0})+2(1-4a)H_{x_2}]H_{x_2x_2x_2}=0,\nonumber\\
&&J_{x_2}-5(2a+1)(2x_1x_2\partial_{x_1x_0}+x_1^2\partial_{x_0}^2+x_2^2\partial_{x_1}^2 +x_2\partial_{x_0})H_{x_2}\nonumber\\
&&\qquad -5(a+2)(x_2\partial_{x_1}+x_1\partial_{x_0})H_{x_1}-5(1+2a)H_{x_0}
-5H_{x_2}(3x_1+2H_{x_1})=0,\nonumber
\end{eqnarray}
\begin{eqnarray}
&&5(a+1)(x_1^3\partial_{x_0}^3+x_2^3\partial_{x_1}^3+3x_2^2x_1\partial_{x_1}^2\partial_{x_0}
+3x_2x_1^2\partial_{x_1}\partial_{x_0}^2+3x_2x_1\partial_{x_0}^2+3x_2^2\partial_{x_1x_0})H
\nonumber\\
&&\qquad+5H_{x_2}(x_1^2\partial_{x_0}^2+x_2^2\partial_{x_1}^2 +2x_2x_1\partial_{x_1x_0}+x_2\partial_{x_0})H
+5(x_1H_{x_1}+3x_1^2+H)H_{x_0}\nonumber\\
&&\qquad+5x_2H_{x_1}(H_{x_1}+3x_1) -x_1J_{x_0}-x_2J_{x_1}-15x_2(3x_1^2-H)=0,\label{DS2}\\
&&-3(x_1^2\partial_{x_0}^2 +x_2^2\partial_{x_1}^2+2x_1x_2\partial_{x_1x_0}+x_2\partial_{x_0})\partial_{x_2}^3H
+(3x_1\partial_{x_0}+3x_2\partial_{x_1}-2H_{x_2})H_{x_1x_2x_2}
+6H_{x_1x_1x_2}\nonumber\\
&&\qquad-10H_{x_2x_2}H_{x_1x_2}+2(3x_1-H_{x_1}+H_{x_2}^2)\partial_{x_2}^3H=0,\nonumber
\end{eqnarray}
\begin{eqnarray}
&&(a-1)^2(9-H_{x_1x_1}+x_1H_{x_0x_1x_2}+x_2H_{x_1x_1x_2}) +(1-a)(2H_{x_2}H_{x_1x_2}+3H_{x_1x_1}+3H_{x_0x_2}-9)
\nonumber\\
&&\qquad+9x_1H_{x_0x_1x_2}+9x_2H_{x_1x_1x_2}-6x_1H_{x_0x_2}H_{x_2x_2} -6H_{x_1x_2}(x_2H_{x_2})_{x_2}\nonumber\\
&&\qquad
+H_{x_2x_2}(9x_1-3H_{x_1}+H_{x_2}^2)=0,\label{DS3}
\end{eqnarray}
\begin{eqnarray}
&& -(a^2+2)(x_1^3 \partial_{x_0}^3+3 x_1^2 x_2 \partial_{x_0}^2\partial_{x_1}+3x_1x_2^2\partial_{x_0}\partial_{x_1}^2 +x_2^3\partial_{x_1}^3 +3x_1x_2\partial_{x_0}^2+3x_2^2\partial_{x_0x_1})H_{x_2} \nonumber\\
&&\qquad +(2a+1)[(x_1^2H_{x_0x_0}+2x_1x_2H_{x_0x_1}+x_2^2H_{x_1x_1}+x_2H_{x_0})H_{x_2x_2}
+(x_1^2H_{x_0x_0x_2}+2x_1x_2H_{x_0x_1x_2}\nonumber\\
&&\qquad +x_2^2H_{x_1x_1x_2}+x_2H_{x_0x_2})H_{x_2} +x_1H_{x_0}H_{x_1x_2}]
+(1-a)[x_1(a-1)H_{x_0x_0}-(a+2)(x_1^2H_{x_0x_0x_1}\nonumber\\
&&\qquad +2x_1x_2H_{x_0x_1x_1}
+x_2^2H_{x_1x_1x_1})+2 x_2H_{x_2}H_{x_1x_1}
+2 x_1H_{x_2}H_{x_0x_1}+2x_1H_{x_1}H_{x_0x_2}-3x_2H_{x_0x_1}]
\nonumber\\
&&\qquad +3(2-a)(x_1H_{x_0x_2}+x_2H_{x_1x_2})^2-x_2(H_{x_2}^2+9x_1-3H_{x_1})H_{x_1x_2} -x_1H_{x_0x_2}H_{x_2}^2-15ax_2H_{x_2}\nonumber\\
&&\qquad +2aHH_{x_0x_2}
-9x_1^2H_{x_0x_2}+6x_2H_{x_2}+HH_{x_0x_2}=0, \label{DS4}
\end{eqnarray}
\begin{eqnarray}
&&4(a-5a^2-5)x_2H_{x_0x_1x_2}+(7-a)(4x_1x_2H_{x_2x_2}H_{x_0x_1x_2} +2 x_1^2 H_{x_2x_2}H_{x_0x_0x_2} +2 x_2^2 H_{x_2x_2}H_{x_1x_1x_2}
\nonumber\\
&&\qquad
+2x_2H_{x_0x_2}H_{x_2x_2})+(4-a)(x_1H_{x_1}H_{x_0x_2x_2} +4x_2H_{x_2x_2}H_{x_1x_1}+4x_1H_{x_2x_2}H_{x_0x_1} +6x_2H_{x_1x_2}^2
\nonumber\\
&&\qquad
+6x_1H_{x_0x_2}H_{x_1x_2})+(1-a)[2H_{x_2}H_{x_1x_1} -2(a+2)H_{x_0x_1}-6(a+1)(x_1\partial_{x_0}\partial_{x_1}^2-x_2\partial_{x_1}^3)H]
\nonumber\\
&&\qquad
-6(a^2+2)(x_2^2H_{x_1x_1x_1x_2}+2x_1x_2H_{x_0x_1x_1x_2} +x_1^2H_{x_0x_0x_1x_2})+(2+a)[(a-2)(x_2^3H_{x_1x_1x_1x_2x_2}
\nonumber\\
&&\qquad
+x_1^3 H_{x_0x_0x_0x_2x_2}+3x_1^2x_2H_{x_0x_0x_1x_2x_2} +3x_1x_2^2H_{x_0x_1x_1x_2x_2}+3x_1x_2H_{x_0x_0x_2x_2}+3x_2^2H_{x_0x_1x_2x_2})
\nonumber\\
&&\qquad
+2x_1H_{x_2}H_{x_0x_1x_2}+2x_2H_{x_2}H_{x_1x_1x_2}+x_1H_{x_0}H_{x_1x_2x_2} +(x_1^2H_{x_0x_0}+x_2^2H_{x_1x_1}+2x_1x_2H_{x_0x_1}
\nonumber\\
&&\qquad
+x_2H_{x_0})H_{x_2x_2x_2}]+2(10-7a)(x_2H_{x_1x_2}+x_1H_{x_0x_2})(x_1H_{x_0x_2x_2} +x_2H_{x_1x_2x_2})
\nonumber\\
&&\qquad
+3aH_{x_2}(x_1^2\partial_{x_0}^2+2x_1x_2\partial_{x_0x_1} +x_2^2\partial_{x_1}^2)H_{x_2x_2}-2x_1(7a^2-2a+4)H_{x_0x_0x_2} -(2x_1H_{x_2}^2-3ax_2H_{x_2}
\nonumber\\
&&\qquad
-aH+18x_1^2-2H)H_{x_0x_2x_2}
-(2x_2H_{x_2}^2-6x_2H_{x_1}+18x_1x_2)H_{x_1x_2x_2} +4(1+2a)H_{x_2}H_{x_0x_2}
\nonumber\\
&&\qquad
-(18x_1-2H_{x_2}^2 -6H_{x_1})H_{x_1x_2}+[(4+5a)H_{x_0}-18ax_2]H_{x_2x_2} -18aH_{x_2}=0, \label{DS5}
\end{eqnarray}
\begin{eqnarray}
&& (1-5a^2-5a)(10x_2^4\partial_{x_0}\partial_{x_1}^3+10x_2x_1^3\partial_{x_0}^4 +30x_1x_2^3\partial_{x_0}^2\partial_{x_1}^2+30x_1^2x_2^2\partial_{x_1}\partial_{x_0}^3 +x_2^5\partial_{x_1}^5+x_1^5\partial_{x_0}^5
\nonumber\\
&&\qquad
+5x_1x_2^4\partial_{x_0}\partial_{x_1}^4 +5x_2x_1^4\partial_{x_0}^4\partial_{x_1} +10x_2^2x_1^3\partial_{x_0}^3\partial_{x_1}^2+10x_1^2x_2^3\partial_{x_0}^2\partial_{x_1}^3
)H
\nonumber\\
&&\qquad
-5(2a+1)(2x_1x_2H_{x_0x_1x_2}+x_1^2H_{x_0x_0x_2}+x_2^2H_{x_1x_1x_2})
(x_1^2H_{x_0x_0}+x_2^2H_{x_1x_1}+2x_2x_1H_{x_0x_1}+x_2H_{x_0})
\nonumber\\
&&\qquad
-5x_1^2(ax_2H_{x_1x_2}+ax_1H_{x_0x_2} +H_{x_2}^2 +2x_2H_{x_1x_2}+2x_1H_{x_0x_2})(x_1\partial_{x_0}^3+3x_2\partial_{x_0}^2\partial_{x_1})H
\nonumber\\
&&\qquad
-5x_2^2
[(a+2)(x_2H_{x_1x_2}+x_1 H_{x_0x_2}) +(2a+1)H_{x_1}+H_{x_2}^2-3(1-a)x_1](x_2\partial_{x_1}^3+3x_1\partial_{x_1}^2\partial_{x_0})H
\nonumber\\
&&\qquad
+5(2a+1)\big\{-H_{x_2}[6x_2^3\partial_{x_0}\partial_{x_1}^2
+x_1^4\partial_{x_0}^4+x_2^4 \partial_{x_1}^4 +6x_1^2x_2\partial_{x_0}^3
+12 x_1x_2^2\partial_{x_0}^2\partial_{x_1}+4x_1x_2^3\partial_{x_0}\partial_{x_1}^3
\nonumber\\
&&\qquad
+6 x_1^2 x_2^2\partial_{x_0}^2\partial_{x_1}^2 +4x_1^3x_2\partial_{x_0}^3\partial_{x_1}] -H_{x_1}[x_1^3\partial_{x_0}^3+3x_2x_1^2\partial_{x_0}^2\partial_{x_1}+3x_2x_1\partial_{x_0}^2 +3x_2^2\partial_{x_0}\partial_{x_1}]-x_2^3 H_{x_0x_2}\partial_{x_1}^2
\nonumber\\
&&\qquad
-H_{x_0}[x_1^2\partial_{x_0}^2+x_2^2\partial_{x_0}\partial_{x_2}+x_2\partial_{x_0}]\big\}H
-5H_{x_0x_0}[(a+2)x_1^3H_{x_0x_1}+x_1x_2(a+2)(3x_2H_{x_1x_2}
\nonumber\\
&&\qquad
+x_1H_{x_1x_1})
+x_2 x_1^2(5a+7)H_{x_0x_2}+x_1H_{x_2}(3x_2H_{x_2}+2x_1H_{x_1}) +3(2ax_2^2+x_1^3+x_2^2)H_{x_2}
\nonumber\\
&&\qquad
-9(1-a)x_2x_1^2] -15[x_1(5a^2x_2^2+ax_1^3+5ax_2^2-x_1^3-x_2^2)\partial_{x_0}^3 +x_2(5a^2x_2^2+3ax_1^3+5ax_2^2-3x_1^3
\nonumber\\
&&\qquad
-x_2^2)\partial_{x_0}^2\partial_{x_1}]H-10(a+2)x_2x_1^2H_{x_0x_1}^2 -5H_{x_0x_1}[3(a+2)(x_2^3H_{x_1x_2}+x_1x_2^2H_{x_1x_1})
\nonumber\\
&&\qquad
+x_1x_2^2(7a+8)H_{x_0x_2}+(5a+4)x_1x_2H_{x_0} +x_2H_{x_2}(3x_2H_{x_2}+4x_1H_{x_1}+6x_1^2)-9(1-a)x_1x_2^2]
\nonumber\\
&&\qquad
-5(a+2)x_2^3H_{x_1x_1}^2+a(3x_2x_1^2\partial_{x_0}^2\partial_{x_1}+3 x_2^2 x_1\partial_{x_0}\partial_{x_1}^2 +x_2^3\partial_{x_1}^3+x_1^3 \partial_{x_0}^3)J-5x_2^2[(2H_{x_1}+3x_1)H_{x_2}
\nonumber\\
&&\qquad
+3(a+1)H_{x_0}]H_{x_1x_1}+x_1(x_1H_{x_2}+3ax_2)J_{x_0x_0}+x_2(2x_1H_{x_2}+3 ax_2)J_{x_0x_1} +x_2^2H_{x_2}J_{x_1x_1}
\nonumber\\
&&\qquad
-(10x_2H_{x_2}H_{x_1}+15x_1x_2H_{x_2} +x_1J_{x_1}-J)H_{x_0}+J_{x_0}(x_2H_{x_2}+x_1H_{x_1}-H)=0,\label{DS6}
\end{eqnarray}
\begin{eqnarray}
&& -(a+2)(x_1^2 H_{x_0x_0}+x_2^2H_{x_1x_1}+2x_2x_1H_{x_0x_1} +x_2H_{x_0})(x_2\partial_{x_1}+x_1\partial_{x_0})H_{x_2x_2}-(a+8)(x_2H_{x_1x_2}
\nonumber\\
&& \qquad
+x_1H_{x_0x_2})(2x_1x_2\partial_{x_0x_1}
+x_1^2 \partial_{x_0}^2+x_2^2 \partial_{x_1}^2 +x_2\partial_{x_0})H_{x_2}-[x_1^2 H_{x_2}^2+(1-a)x_1^2H_{x_1}
\nonumber\\
&& \qquad
+3ax_1x_2H_{x_2} +x_1(a+2)H-3(2a^2+1)x_2^2-9x_1^3]H_{x_0x_0x_2} -[(a+2)x_1^2H_{x_0}+(4-a)x_1x_2H_{x_1}\nonumber\\
&& \qquad
+2x_1x_2H_{x_2}^2 +3ax_2^2H_{x_2}+x_2(aH-18x_1^2+2H)]H_{x_0x_1x_2}+(2a^2+1)[6x_1^2x_2\partial_{x_0}^3 +6x_2^3\partial_{x_0}\partial_{x_1}^2
\nonumber\\
&& \qquad
+12x_1x_2^2\partial_{x_0}^2\partial_{x_1}+x_1^4\partial_{x_0}^4 +4x_1x_2^3\partial_{x_0}\partial_{x_1}^3+6x_1^2x_2^2\partial_{x_0}^2\partial_{x_1}^2 +4x_2 x_1^3\partial_{x_0}^3\partial_{x_1}+x_2^4\partial_{x_1}^4]H_{x_2}
\nonumber\\
&& \qquad
+(1-a)[(a+1)(6 x_1 x_2^2\partial_{x_1}^3\partial_{x_0}+6x_1^2x_2\partial_{x_0}^2\partial_{x_1}^2 +2x_1^3\partial_{x_0}^3\partial_{x_1}+2x_2^3\partial_{x_1}^4)H
\nonumber\\
&& \qquad
-H_{x_2}(x_2^2\partial_{x_1}^3
+2x_1x_2\partial_{x_0}\partial_{x_1}^2+x_1^2\partial_{x_0}^2\partial_{x_1}+2x_1\partial_{x_0}^2
+3x_2\partial_{x_0x_1})H-x_1H_{x_0}H_{x_1x_1}]
\nonumber\\
&& \qquad
-(10-7a)x_1H_{x_0x_2}(x_2H_{x_1x_1}+x_1H_{x_0x_1})+2x_1^2 H_{x_2}H_{x_0x_2}^2
+[9(2a+1)x_1x_2+4 x_2 x_1H_{x_2}H_{x_1x_2}
\nonumber\\
&& \qquad
-(5a+4)x_1H_{x_0}+(2x_1H_{x_2}+3x_2)H_{x_1}-x_2H_{x_2}^2 -2H_{x_2}H]H_{x_0x_2}
-3 ax_1x_2^2H_{x_2}H_{x_0x_1x_1x_2}
\nonumber\\
&& \qquad
+(a+2)x_2^2(4-3x_1H_{x_2x_2}-a)H_{x_0x_1x_1} -[x_2(4x_1H_{x_2}+3(a+2)x_2)H_{x_2x_2}+(16-7a)x_1x_2H_{x_1x_2}
\nonumber\\
&& \qquad
+(a+2)x_1H_{x_1}-x_1H_{x_2}^2-(aH+9x_1^2-H)]H_{x_0x_1}
+x_2H_{x_2}^2(6-x_2H_{x_1x_1x_2}+H_{x_1x_1})
\nonumber\\
&& \qquad
+H_{x_2}(2x_2^2H_{x_1x_2}\partial_{x_1}-ax_1^3\partial_{x_0}^3
-ax_2^3\partial_{x_1}^3-3ax_2x_1^2\partial_{x_0}^2\partial_{x_1}
-2x_1H_{x_0}\partial_{x_1})H_{x_2}-H_{x_2x_2}[2(x_1^2 H_{x_0x_0}
\nonumber\\
&& \qquad
+x_2^2H_{x_1x_1}+x_2H_{x_0})H_{x_2} +(a+2)(x_1^3\partial_{x_0}^3+x_2^3\partial_{x_1}^3+3x_2x_1^2\partial_{x_0}^2\partial_{x_1}
+3x_2x_1\partial_{x_2}^2)H]
\nonumber\\
&& \qquad
+(5a^2+2a+2)x_1^2H_{x_0x_0x_0}+2(2a^2+2a+5)x_1x_2H_{x_0x_0x_1} -[(a+2)x_1x_2H_{x_0}+3x_2^2H_{x_1}
\nonumber\\
&& \qquad
-9x_2^2x_1]H_{x_1x_1x_2}+[(5a^2+2a+2)x_2-3x_1^2H_{x_1x_2}]H_{x_0x_0} +[(7a-13)x_2^2H_{x_1x_2}-3x_2H_{x_1}
\nonumber\\
&& \qquad
+9x_1x_2]H_{x_1x_1}-54ax_2x_1+3[(5a+1)x_1+2aH_{x_1}]H_{x_0}+18ax_2H_{x_1} -[(4a+5)x_2H_{x_0}
\nonumber\\
&& \qquad-18ax_2^2]H_{x_1x_2}-3x_1x_2(a+2)H_{x_0x_0}H_{x_2x_2}-3aJ_{x_0}/5=0. \label{DS7}
\end{eqnarray}
From the determining equation \eqref{DS}, we know that the determining equations \eqref{DS}-\eqref{DS7} should be solved in three separated cases for $a=0,\ a=1$ and $a\neq 0,\ 1$, respectively. Now, it is easy to finish the final work by solving \eqref{DS}-\eqref{DS7}.\\
\bf \it Case 1. $a=0$. \rm In this simple case, the final solutions of $H$ and $J$ read
\begin{equation}\label{HJ_SK}
H=cx_1+c_1=cw_x+c_1,\
J=15cw_x^2-15w_x^3+(5c^2+15c_1)w_x.
\end{equation}
\bf \it Case 2. $a=1$. \rm In this case, one can find four different solutions
\begin{equation}\label{HJ_KdV1}
H=\frac32x_1^2-c_1,\
J=\frac{45}2(x_2^2+x_1^3)+15c_1x_1,\ x_0=w,\ x_1=w_x,\ x_2=w_{x2},
\end{equation}
\begin{equation}\label{HJ_KdV2}
H = cw_x+3w_x^2+c_1,\ J = 90w_x^3+45cw_x^2+5c^2w_x
+15c_1w_x+45w_{x2}^2,
\end{equation}
\begin{equation}\label{HJ_JM}
H = c_1-\frac{3w_{x2}^2}{4(c+w_x)}+\frac32w_x^2,\
J = 15c_1w_x-\frac{315w_{x2}^4}{16(c+w_x)^3}-\frac{45(w_x-2c)}{4(c+w_x)}w_{x2}^2 +\frac{45}2w_x^3,
\end{equation}
\begin{eqnarray}\label{HJ_STO}
H&=&\frac34[2(w+c)w_{x2}+2w_x^2+(c+w)^2w_x],\\
J&=&\frac{45}{16}\{16w_{x2}^2+[40(w+c)w_x+4(w+c)^3]w_{x2} +12w_x^3+12(w+c)^2w_x^2+(w+c)^4w_x\}.\nonumber
\end{eqnarray}
\bf \it Case 3. $a\neq0,\ 1$. \rm In this case, we find that $a$ should be fixed as $a=-\frac12$ and the functions $H$ and $J$ are also be fixed as
\begin{equation}\label{HJ_KdV3}
H=-\frac32w_x^2+cw_x+c_1, J = -\frac{45}4w_{xx}^2-\frac{45}2w_x^3+5c^2w_x+15c_1w_x.
\end{equation}
$c$ and $c_1$ in all cases are arbitrary constants.

By substituting these solutions \eqref{HJ_SK}--\eqref{HJ_KdV3} into the decomposition relations \eqref{wy1} and \eqref{wt1}, proposition 2 is proved.

Proposition 2 provides simple and interesting relationships between classical integrable systems and PBKP equation \eqref{PBKP}.
The first three decompositions \eqref{DC51}--\eqref{DC53} are all the KdV decompositions. The first equation of the fourth decomposition \eqref{DC5JM} is an integrable model which is a special form of the Svinolupov-Sokolov (SS) equation proposed in \cite{SS}. The decomposition \eqref{DCSTO} is related to the so-called Sharma-Tasso-Olver (STO) equation (the third-order equation of the Burgers hierarchy) \cite{STO1,STO2}. The sixth decomposition is a natural SK decomposition. The unexpected linear superposition property of these decompositions continues to surprise us in Sec. \ref{slspbkp}.

\subsection{Decompositions of the seventh-order PBKP equation (\ref{K7})}

For the seventh-order PBKP equation, decompositions can be found in a similar way. We just list the results in the following proposition.\\
\bf Proposition 3. \rm The functions $w_i,\ i=1,\ 2,\ \ldots,\ 6$ satisfying the following decomposition systems
\begin{eqnarray}
&&\left\{\begin{array}{l}
\displaystyle{w_{1y}=(\Phi_1+c) w_{1x}+c_1},\\
\\
\displaystyle{w_{1t}=[27\Phi_1^3+63c \Phi_1^2 +(42c^2+63c_1)\Phi_1+7c(c^2+9c_1)]w_{1x}},
\end{array}\right. \label{DC71}\\
&&\left\{\begin{array}{l}
\displaystyle{w_{2y}=\Phi_2w_{2x}+c_1},\\
\\
\displaystyle{w_{2t}=(27\Phi_2^3+63c_1\Phi_2)w_{2x}},
\end{array}\right. \label{DC72}\\
&&\left\{\begin{array}{l}
\displaystyle{w_{3y}=-\frac12( \Phi_3-2c)w_{3x}+c_1},\\
\\
\displaystyle{w_{3t}=\bigg[\frac{27}8 \Phi_3^3-\frac{63}4 c \Phi_3^2+\frac72(3c^2-9c_1)\Phi_3+7c(c^2+9c_1)\bigg]w_{3x}},
\end{array}\right. \label{DC73}\\
&&\left\{\begin{array}{l}
\displaystyle{w_{4y}=\Phi_4w_{4x}+c_1},\ \\
\\
\displaystyle{{w_{4t}}=27\Phi_4^3w_{4x}+63c_1\Phi_4w_{4x}},
\end{array}\right. \label{DC74}\\
&&\left\{\begin{array}{l}
\displaystyle{w_{5y}=\frac14(4\Phi_5^2+6c\Phi_5+3c^2)w_{5x}},\ ,\\
\\
\displaystyle{w_{5t}=\frac{27}{64}(64\Phi_5^6+224c\Phi_5^5 +336c^2\Phi_5^4
	+280 c^3\Phi_5^3+140c^4\Phi_5^2+42c^5\Phi_5+7c^6)w_{5x}},
\end{array}\right. \label{DC75}\\
&&\left\{\begin{array}{l}
\displaystyle{w_{6y}=cw_{6x}+c_1},\\
\\
\displaystyle{w_{6t}=84c_1cw_{6x}+7c^3w_{6x}+21(3w_{6x}^2+w_{6x3})c^2-21(w_{6x}^3 +w_{6x}w_{6x3}-w_{6x2}^2)c }\\
\\
\displaystyle{\qquad\quad -63w_{6x}(w_{6x}^3+2w_{6x}w_{6x3}+w_{6x2}^2)-21(w_{6x}w_{6x5}+w_{6x2}w_{6x4}+w_{6x3}^2)-w_{6x7}}
\end{array}\right. \label{DC76}
\end{eqnarray}
are all solutions of the seventh-order PBKP equation \eqref{K7} with $t_7=t$.

To prove proposition 3, one can directly substitute the decompositions to the seventh-order PBKP equation \eqref{K7} and finish some integrations by parts. Here, we omit the details on the proof procedure. 
Same as in the fifth-order BKP equation, decompositions make it possible to find interrelations between classical integrable systems and the seventh-order PBKP equation.
The first three decompositions are the usual potential KdV and seventh-order potential KdV reductions. The fourth decomposition is related to the special SS equation \eqref{DC5JM} and its seventh-order flow. The fifth decomposition is related to the third-order Burgers (STO) and seventh-order Burgers equations in potential forms. The sixth decomposition is simply the seventh-order potential SK reduction of the seventh-order PBKP equation. 

\subsection{Decompositions of the ninth-order PBKP equation (\ref{K9})}

In the same way, we directly write down the decomposition proposition for the ninth-order PBKP equation \eqref{K9} without detailed verifications. \\
\bf Proposition 4. \rm The functions $w_i,\ i=1,\ \ldots,\ 6$ provide solutions of the ninth-order PBKP equation \eqref{K9} with $t_9=t$ if they satisfy
\begin{eqnarray}
&&\left\{\begin{array}{l}
\displaystyle{w_{1y}=(\Phi_1+c) w_{1x}+c_1},\quad \mu\equiv \frac92(2c^4+54c_1c^2+45c_1^2),\\
\\
\displaystyle{w_{1t}=[81\Phi_1^4+243c\Phi_1^3+243(c^2+c_1)\Phi_1^2
	+18c(5c^2+27c_1) \Phi_1 +\mu] w_{1x}},
\end{array}\right. \label{DC91}\\
&&\left\{\begin{array}{l}
\displaystyle{w_{2y}=\Phi_2w_{2x}+c_1},\\
\\
\displaystyle{w_{2t}=\bigg(81\Phi_2^4+243c_1\Phi_2^2+\frac{405}2c_1^2\bigg)w_{2x}-c},
\end{array}\right. \label{DC92}\\
&&\left\{\begin{array}{l}
\displaystyle{w_{3y}=-\frac12(\Phi_3-2c) w_{3x}+c_1},\\
\\
\displaystyle{w_{3t}=\bigg[\frac{81}{16}\Phi_3^4-\frac{81(2c^2+3c_1)}{4} \Phi_3^2+\frac{9c(8c^2-9c_1)}{2} \Phi_3+\mu\bigg]w_{3x}},
\end{array}\right. \label{DC93}\\
&&\left\{\begin{array}{l}
\displaystyle{w_{4y}=\Phi_4w_{4x}+c_1,\ }\\
\\
\displaystyle{{w_{4t}}=81\Phi_4^4w_{4x}+243c_1\Phi_4^2w_{4x}+\frac{405c_1^2}{2}w_{4x}},
\end{array}\right. \label{DC94}\\
&&\left\{
\begin{array}{l}
\displaystyle{w_{5y}=\frac14(4\Phi_5^2+6c\Phi_5+3c^2)w_{5x},}\\
\\
\displaystyle{w_{5t}=\frac{81}{256}(256\Phi_6^8+1152c\Phi_5^7+2304c^2\Phi_5^6+2688c^3\Phi_5^5
	+2016c^4\Phi_5^4 +1008c^5\Phi_5^3}\\ \\
\displaystyle{\qquad\quad +336c^6\Phi_5^2+72c^7\Phi_5+9c^8)w_{5x}},
\end{array}\right. \label{DC95}\\
&&\left\{\begin{array}{l}
\displaystyle{w_{6y}=cw_{6x}+c_1},\\
\\
\displaystyle{w_{6t}=-9(63w_{6x}^4+126w_{6x}^2w_{6x3}+63w_{6x}w_{6x2}^2+21w_{6x}w_{6x5}
	+21w_{6x2}w_{6x4}+21w_{6x3}^2+w_{6x7})c }\\ \\
\displaystyle{\qquad\quad +54(3w_{6x}^2+w_{6x3})c^3+27(8w_{6x}^3+8w_{6x}w_{6x3}+8c_1w_{6x}+7w_{6x2}^2+w_{6x5})c^2 +135(3w_{6x}^2}\\ \\
\displaystyle{\qquad\quad +w_{6x3})c_1c +\frac92w_{6x}(2c^4+45c_1^2)-27(15w_{6x}^3+15w_{6x}w_{6x3}+w_{6x5})c_1}.
\end{array}\right. \label{DC96}
\end{eqnarray}

Similar to the decompositions of the fifth- and seventh-order PBKP equations, the first three decompositions \eqref{DC91}--\eqref{DC93} are the potential KdV decompositions, the fourth decomposition \eqref{DC94} is the special SS decomposition, the fifth decomposition \eqref{DC95} is the higher-order Burgers decomposition and the sixth decomposition \eqref{DC96} is the SK reduction.
By observing the decomposition propositions 2, 3, and 4, we conclude that classical integrable systems solve the PBKP hierarchy after simple decompositions.
\\
\bf Conjecture 2. \rm  Each  order equation in PBKP hierarchy \eqref{PBKPH}  possesses six types of $\{y,\ t\}$ decomposed solutions. 

The first three types of KdV decompositions can be obtained from the nonlinearization procedure of the Lax pairs of the PBKP hierarchy, the fourth decomposition is the special SS decomposition and the fifth type of the STO decomposition can be concluded from hints of the B\"acklund transformations of the bilinear PKP equation. The sixth type of SK reduction is a direct conclusion because the PBKP hierarchy is just a (2+1)-dimensional extension of the (1+1)-dimensional SK hierarchy.

\section{Special linear superpositions of the PBKP hierarchy}\label{slspbkp}
It is well known that for the physical systems frequently characterized by nonlinear differential equations, there is no linear superposition theorem.
However, there exist some types of nonlinear superposition properties if the solutions are linked by some special requirements such as the B\"aclund/Darboux transformations \cite{Hirota,BGK,Hu96,NSP1,HuB,HZ,MM}.
For the BKP equation (alias (2+1)-dimensional Sawada-Kotera equation), the nonlinear superpositions have been studied by Li and Hu \cite{LiHu} in bilinear forms. In this section, we investigate the possible linear superpositions of the PBKP hierarchy for some special decomposition solutions obtained in the previous section.

\subsection{Special linear superpositions of the fifth-order PBKP equation (\ref{PBKP})}

For the fifth-order PBKP equation \eqref{PBKP}, if $w_i$ and $w_j$ are the decompositions of the PBKP equation listed in the previous section, then requiring the linear combination $a_1w_i+a_2w_j$ is also a solution of the PBKP equation \eqref{PBKP} yields the following special linear superposition proposition. 
\\
\bf Proposition 5. \rm Suppose that $w_1,\ w_2,\ w_3,\ w_4,\ w_5$ and $w_6$ are solutions of the PBKP equation \eqref{PBKP} with the conditions
\begin{eqnarray}\begin{split}\label{w16}
&w_{1y}=(\Phi_1+c) w_{1x},\
w_{1t}=(9\Phi_1^2+15c \Phi_1+5c^2)w_{1x},\ \Phi_1\equiv \partial_x^2+4w_{1x}-2\partial_x^{-1}w_{1x2}, \\
&w_{2y}=(\Phi_2-c) w_{2x},\
w_{2t}=(9\Phi_2^2-15c \Phi_2+5c^2)w_{2x}, \ \Phi_2\equiv \partial_x^2+4w_{2x}-2\partial_x^{-1}w_{2x2},\\
&w_{3y}=\Phi_3 w_{3x},\
w_{3t}=9\Phi_3^2w_{3x}, \ \Phi_3\equiv \partial_x^2+4w_{3x}-2\partial_x^{-1}w_{3x2}, \\
&w_{4y}=\Phi_4w_{4x},\ w_{4t}=9\Phi_4^2w_{4x}, \ \Phi_4\equiv \partial_x^2+2w_{4x}-\partial_x^{-1}w_{4x2},\\
&w_{5y}=\Phi_5w_{5x}-\frac{c^2}6,\ w_{5t}=(9\Phi_5^2-\frac{5c^2}2)w_{5x}, \ \Phi_5\equiv \partial_x^2+2w_{5x}-\partial_x^{-1}w_{5x2},\\
&w_{6y}=\Phi_6w_{6x}-\frac{c^2}6,\ w_{6t}=(9\Phi_6^2-\frac{5c^2}2)w_{6x}, \ \Phi_6\equiv \partial_x^2+2w_{6x}-\partial_x^{-1}w_{6x2},
\end{split}\end{eqnarray}
then the linear superpositions 
\begin{eqnarray}\label{w79}
&&w_7=w_1+w_2,\ w_8=w_3+\frac12w_4,\ w_9=\frac12(w_5+w_6)
\end{eqnarray} 
are at the same time the solutions of the PBKP equation \eqref{PBKP}. \\
\bf Proof. \rm By substituting the linear superposition relations \eqref{w79} with the decomposition results \eqref{w16} into the PBKP equation \eqref{PBKP}, one can directly prove the proposition.

In fact, one can directly substitute the general superposition assumption $w=f(w_i,\ w_j)$ with $w_i$ and $w_j$ being any one of the decompositions given in the proposition 2 to prove that the results \eqref{w79} are the only possible superpositions if there is no further relations among $w_i$ and $w_j$.

\begin{figure}[h]
	\subfigure[]{\includegraphics[width=0.32\textwidth]{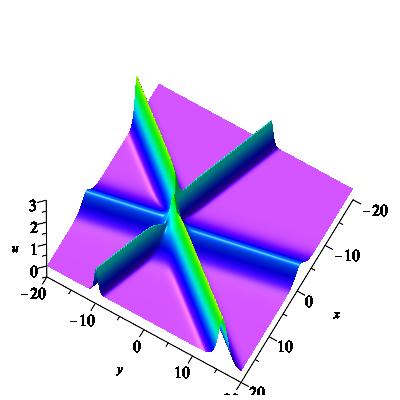}} \subfigure[]{\includegraphics[width=0.32\textwidth]{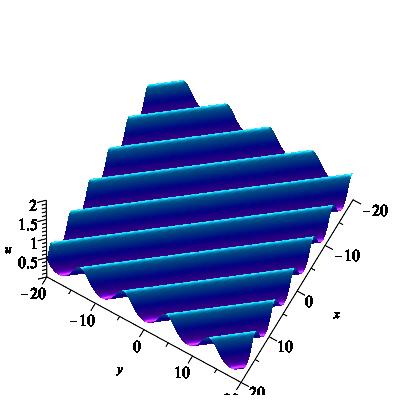}}
	\subfigure[]{\includegraphics[width=0.32\textwidth]{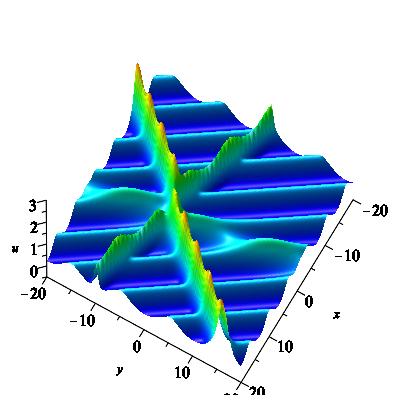}}
	\caption{Structures of three-soliton (a), periodic cnoidal wave (b) and their linear superposition (c) of BKP equation at $t=0$, respectively.}
	\label{fig:case2}       
\end{figure}

\begin{figure*}[h]
	\subfigure[]{\includegraphics[width=0.32\textwidth]{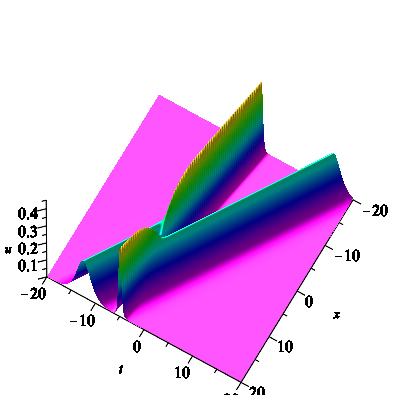}} \subfigure[]{\includegraphics[width=0.32\textwidth]{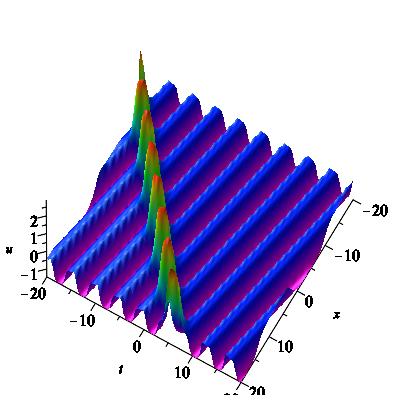}}
	\subfigure[]{\includegraphics[width=0.32\textwidth]{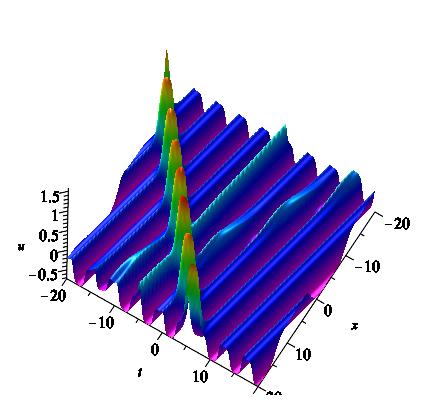}}
	\caption{Two-soliton (a), soliton-cnoidal interaction wave (b) and their linear combination (c) of BKP equation at $y=0$, respectively. }
	\label{fig:case3}       
\end{figure*}

\begin{figure*}[h]
	\subfigure[]{\includegraphics[width=0.32\textwidth]{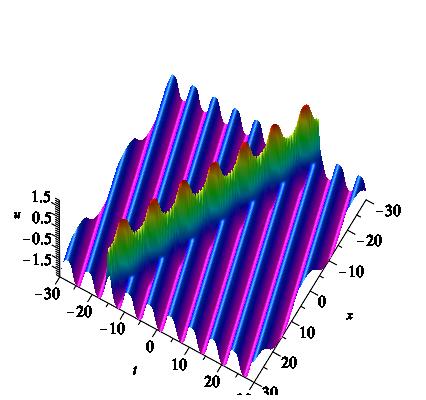}} \subfigure[]{\includegraphics[width=0.32\textwidth]{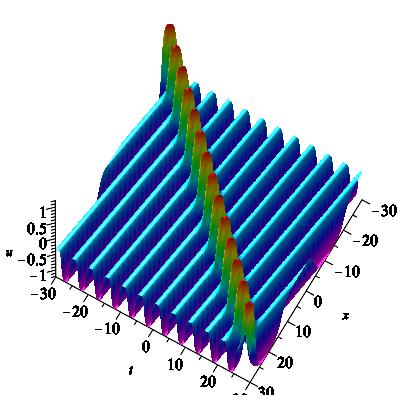}}
	\subfigure[]{\includegraphics[width=0.32\textwidth]{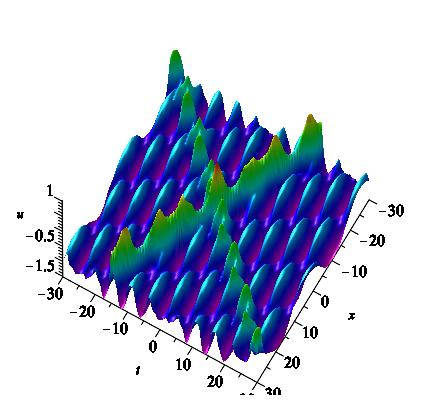}}
	\caption{Two soliton-cnoidal interaction waves (a) and (b), their linear superposition (c) of BKP equation at $y=0$.}
	\label{fig:case4}       
\end{figure*}

It is noted that though there are six possible decompositions for the fifth-order PBKP equation \eqref{PBKP} as shown in the proposition 2, only the first two types of decomposition solutions \eqref{DC51} and \eqref{DC52} can be linearly combined to construct new solutions of the PBKP equation \eqref{PBKP}. The third type of KdV decomposition \eqref{DC53}, the special SS decomposition \eqref{DC5JM}, the STO decomposition \eqref{DCSTO} and the third SK reduction \eqref{DCSK} can not be used to find new solutions via superposition assumption $w=f(w_i,\ w_j)$ if there is no further relations among seed solutions $w_i$ and $w_j$.
It should be emphasized that $w_i$ and $w_j,\ i,\ j=1,\ 2,\ \ldots, \ 6,$ appeared in superpositions \eqref{w79} are independent solutions of the decompositions. 
Here, we list some special linear superposition structures of the fifth BKP equation.\\
\bf Example 1. \it $n+m$ solitons. \rm
\begin{eqnarray}
&&w_{1}=2(\ln f_n^+)_{x},  w_{2}=2(\ln f_m^-)_{x}, w_3=\left.w_1\right|_{c=0}, w_4=\left.w_6\right|_{c=0},  w_{5}=4(\ln g_n^+)_{x}+\frac{cx}3, w_{6}=\left.w_5\right|_{g_n^+\rightarrow g_m^-},\label{ns}\\
&&f_{n}^{\pm}=\sum_{\mu=0,1}\exp\left(\sum_{i=1}^n \mu_i\xi_i^{\pm}+\sum_{1\leq i<j\leq n}\mu_i\mu_j\theta_{ij}^{\pm}\right),\ g_{n}^{\pm}=\sum_{\mu=0,1}\exp\left(\sum_{i=1}^n \mu_i\eta_i^{\pm}+\sum_{1\leq i<j\leq n}\mu_i\mu_j\theta_{ij}^{\pm}\right),\label{fngn}
\end{eqnarray}
where the summation of $\mu$ should be done for all permutations of $\mu_i=0,\ 1,\ i=1,\ 2,\ \ldots,\ n$,
\begin{eqnarray}
&&\xi_{i\pm}=k_{i\pm}x+k_{i\pm}(k_{i\pm}\pm c)y+(9k_{i\pm}^5\pm 15ck_{i\pm}^3+5c^2k_{i\pm})t+\xi_{i0}^{\pm},\label{xieta} \\
&&\eta_{i\pm}=k_{i\pm}x+k_{i\pm}(k_{i\pm}+ c)y+(9k_{i\pm}^5+ 15ck_{i\pm}^3+5c^2k_{i\pm})t+\eta_{i0}^{\pm},\ \exp(\theta_{ij}^{\pm})=\frac{(k_{i\pm}-k_{j\pm})^2}{(k_{i\pm}+k_{j\pm})^2}, \nonumber
\end{eqnarray}
and $k_{i\pm},\ c,\ \xi_{i0}^{\pm}$ and $\eta_{i0}^{\pm}$ are arbitrary constants.
All the related solutions of the BKP equation \eqref{BKP} with $u=w_x$ and $w=w_7,\ w_8,\ w_9$ expressed in proposition 5 with \eqref{ns}, \eqref{fngn} and \eqref{xieta} denote some types of $n+m$ solitons solutions. To illustrate the linear superposition structure more clearly, let us look at some figures. The following figures are intended to illuminate two-soliton, three-soliton and their linear superposition of the BKP equation \eqref{BKP} respectively with condition $\{c=2,k_1=-1,k_2=1.5,k_3=-2,\xi_{10}=2,\xi_{20}=10,\xi_{30}=10\}$ at time $t=0$.
\begin{figure*}[h]
	\subfigure[]{\includegraphics[width=0.32\textwidth]{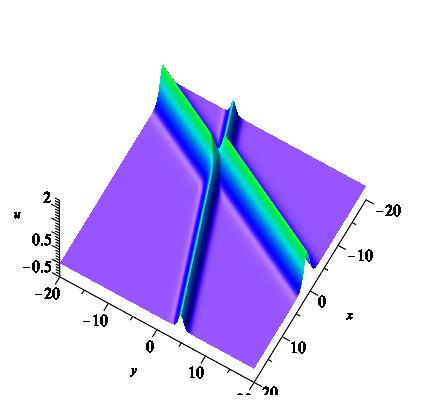}} \subfigure[]{\includegraphics[width=0.32\textwidth]{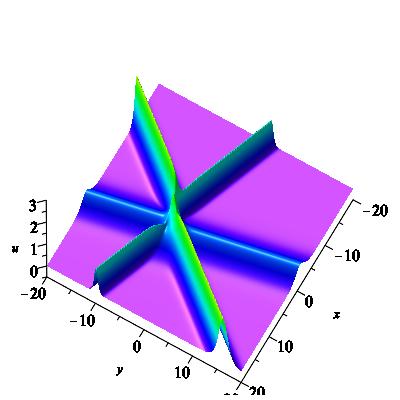}}
	\subfigure[]{\includegraphics[width=0.32\textwidth]{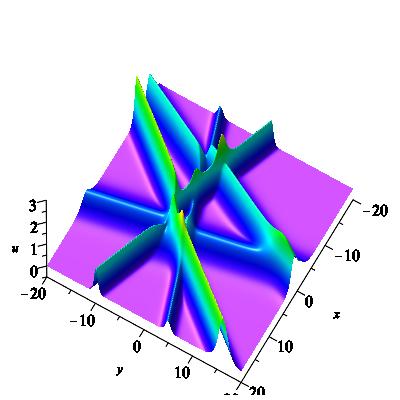}}
	\caption{Two-soliton (a), three-soliton (b) and their linear superposition (c) of BKP equation located at $t=0$, respectively.}
	\label{fig:case1}       
\end{figure*}
\\
\bf Example 2. \it $n$ solitons with periodic cnoidal background waves. \rm
If one of $w_1$ and $w_2$ in example 1 is replaced by a periodic solution, say,
\begin{equation}
w_{2x}=2k^2m^2{\mbox{cn}}^2\big(kx+[4k^3(2m^2-1)-ck]y +[5 c^2 k-60 k^3 (2m^2-1)c+72 k^5 (7m^4-7m^2+2)]t+\xi_0,\ m\big),
\end{equation}
where ${\mbox{cn}}(\xi,\ m)$ is a Jacobi cn function with the modula $m$,  then $w_7=w_1+w_2$ becomes an $n$-soliton solution on a periodic cnoidal background wave. Figure \ref{fig:case2} is drawn for the case of three-soliton moving on a periodic cnoidal wave by taking simply $\{c=1.2,k_1=-1,k_2=1.5,k_3=-2,\xi_0=\xi_{10}=2,\xi_{20}=\xi_{30}=10,m=1.2,k=0.5\}$ at $t=0$.
\\ 
\bf Example 3. \it $n$ solitons solution combined with soliton-cnoidal wave interaction solution. \rm If $w_3$ is given by \eqref{ns} and $w_4$ is fixed as
\begin{eqnarray}
&&w_{4x}=\frac{8}3s_x^2-\frac{4s_{x3}}{3s_x}+\frac{s_{x2}^2}{3s_x^2}+4s_{x2}\tanh(s) -4s_x^2\tanh^2(s),\label{w4}\\
&&s=mkx+l_0y+\omega_0t+\xi_{0}-{\rm arctanh}\big(m{\rm sn}(kx+ly+\omega t+\xi_1,m^2) \big)
\end{eqnarray}
with arbitrary constants $k,\ m,\ \xi_0,\ \xi_1,\ \lambda$ and $l_0=mk\lambda-\frac{mk^3}2(1+3m^2)(m^2+3)$,
$l=k\lambda +\frac{k^3}2\big[(m^2-1)^2-16m^2\big]$, $\omega=15l\lambda-\frac{15}2k\lambda^2+\frac{9k^5}8\big(3m^8+20m^6+722m^4+20m^2+3\big)$ and
$\omega_0=15l_0\lambda -\frac{15}2mk\lambda^2-\frac{9mk^5}8\big(5m^8-180m^6-428m^4-180m^2+5\big)$,
then $w=w_3+\frac12 w_4$ expresses an $n$-soliton solution combined with a soliton-cnoidal wave interaction solution. In figure \ref{fig:case3}, we draw the behavior of a two-soliton, a soliton-cnoidal wave and their linear combination  of BKP equation \eqref{BKP} at $y=0$ with arbitrary parameters fixed as  $\{c=0,k_1=0.9,k_2=0.65,\xi_{10}=12,\xi_{20}=0,k=0.3,m=0.8,\xi_0=-10,\xi_1=10,\lambda=0.3\}$.
\\
\bf Example 4.\ \it Combination solutions of two different modified Schwarzian KdV waves. \rm If the solutions of \eqref{w16} for
$w_5$ and $w_6$ are rewritten in the forms
\begin{eqnarray}
&&w_{5x}=2s_{1x}^2-\frac{s_{1x2}^2}{s_{1x}^2}-S_1+\frac{\lambda_1}3+4s_{1x2}\tanh(s_1) -4s_{1x}^2\tanh^2(s_1),\label{w5}\\
&&w_{6x}=2s_{2x}^2-\frac{s_{2x2}^2}{s_{2x}^2}-S_2+\frac{\lambda_2}3+4s_{2x2}\tanh(s_2) -4s_{2x}^2\tanh^2(s_2),\label{w6}
\end{eqnarray}
where $s_1$ and $s_2$ are any solutions of the modified Schwarzian KdV systems
\begin{eqnarray}
&&Y_{i}=S_i-2s_{ix}^2+\lambda_i,\ Y_i\equiv \frac{s_{iy}}{s_{ix}},\ T_i\equiv \frac{s_{it}}{s_{ix}},\ S_i\equiv \frac{s_{ix3}}{s_{ix}}-\frac{3s_{ix2}^2}{2s_{ix}^2},\ i=1,\ 2,\label{Yi}\\
&&T_i=9S_{ix2}+\frac{27}2S_i^2+15\lambda_iS_i+\frac{5c^2}2+\frac{15\lambda_i^2}2 +6s_{ix}^2(9s_{ix}^2-15S_i-5\lambda_i)-90s_{ixx}^2\label{Ti}
\end{eqnarray}
with arbitrary constants $\lambda_i$, then $w=\frac12(w_5+w_6)$  is a combination solution of two modified Schwarzian KdV waves. The modified Schwarzian KdV systems \eqref{Yi} and \eqref{Ti} are related to the Schwarzian KdV systems (the systems of \eqref{Yi} and \eqref{Ti} contain only the M\"obius transformation invariant terms related to the M\"obius invariants $T_i,\ Y_i,\ S_i,\ S_{ix2},\ \lambda_i$ and $c$)  by the transformation $s_i\rightarrow {\rm arctanh}(s_i^{-1})$.
The modified Schwarzian KdV systems \eqref{Yi} and \eqref{Ti} possess various interaction solutions including the following soliton-cnoidal interaction solutions \cite{CRE,CRE1}
\begin{eqnarray}
&&s_1=m_1k_1x+p_1y+\omega_1t+\xi_{10}-{\rm arctanh}\big[m_1{\rm sn}(k_1x+q_1y+\Omega_1t+\eta_{10},m_1^2)\big],\label{s1}\\
&&s_2=k_0(x+p_2y+\omega_2t)+\xi_{20}-{\rm arctanh}\big[m_2{\rm sn}(k_2x+q_2y+\Omega_2t+\eta_{20},m_2^2)\big]\label{s2}
\end{eqnarray}
with arbitrary constants $k_i,\ m_i,\ \lambda_i,\ \xi_{i0},\ \eta_{i0},\ i=1,\ 2$, and the other constants $\{k_0, p_1,\ q_i,\ \omega_i,\ \Omega_i\}$ being given by
\begin{eqnarray*}
	&&p_1=m_1k_1\lambda_1-\frac{m_1k_1^3}2(m_1^2+3)(3m_1^2+1),\\
	&&q_1=k_1\lambda_1+\frac{k_1^3}2\big[(m_1^2-1)^2-16m_1^2\big],\\
	&&\omega_1=15p_1\lambda_1+\frac52 m_1k_1(c^2-3\lambda_1^2)-\frac98m_1k_1^5\big[5m_1^8-180m_1^2(1+m_1^4) -418m_1^4\big],\\
	&&\Omega_1=15q_1\lambda_1+\frac{5k_1}2(c^2-3\lambda_1^2) +\frac{9k_1^5}8(3m_1^8+20m_1^6+722m_1^4+20m_1^2+3),\\
	&&k_0=-\frac{k_2}2(m_2^2+1),\ p_2=\lambda_2-\frac{k_2^2}2(1+14m_2^2+m_2^4),\\ &&q_2=k_2\lambda_2-\frac{k_2^3}2(5+6m_2^2+5m_2^4),\\
	&&\omega_2=15p_2\lambda_2+\frac{5}2(c^2-2\lambda_2^2) +\frac{9k_2^4}8\big(3m_2^8+212m_2^2(1+m_2^4)+3\big),\\ &&\Omega_2=15q_2\lambda_2-\frac52k_2(c^2-3\lambda_2^2) +\frac{9k_2^5}8\big(43(1+m_2^8)+180m_2^2(1+m_2^4)+322m_2^4\big).
\end{eqnarray*}
Figure \ref{fig:case4} shows a linear superposition structure of two soliton-cnoidal interaction waves for the field $u$  with the conditions $k_1=0.3,k_2=0.25, m_1=1.2,m_2=0.6, \xi_{10}=\xi_{20}=0, \eta_{10}=\eta_{20}=10,\lambda_1=\lambda_2=0$ at $y=0$.
Obviously, figures (c) in figures \ref{fig:case1}-\ref{fig:case4} clearly display  the linear superpositions of (a) and (b), correspondingly. 
Interestingly, two waves can pass through each other without altering each other. Sometimes, this leads to some truly weird behaviors. It also means that waves can constructively or destructively interfere under the effect of linear superposition.

\subsection{Special linear superpositions of the seventh-order PBKP equation (\ref{K7})}
Similar to the fifth-order PBKP equation, though there are six decomposition solutions \eqref{DC71}--\eqref{DC76}, only the first two decompositions \eqref{DC71} and \eqref{DC72} can be used to construct new solutions of the seventh-order PBKP equation (\ref{K7}) via linear superpositions. The result is summarized in the following proposition.\\
\bf Proposition 6. \rm If $w_1,\ w_2,\ w_3,\ w_4,\ w_5$ and $w_6$ are solutions of the seventh-order PBKP equation \eqref{K7} with conditions
\begin{eqnarray}\begin{split}\label{w7pbkp}
&w_{1y}=(\Phi_1+c) w_{1x},\ w_{1t}=(27\Phi_1^3+63c \Phi_1^2 +42c^2\Phi_1+7c^3)w_{1x}-c,\\
&w_{2y}=(\Phi_2-c) w_{2x},\ w_{2t}=(27\Phi_2^3-63c \Phi_2^2 +42c^2\Phi_2-7c^3)w_{2x}-c_1,\\
&w_{3y}=\Phi_3 w_{3x},\ w_{3t}=27\Phi_3^3w_{3x}-c,\\
&w_{4y}=\Phi_4w_{4x},\ w_{4t}=27\Phi_4^3w_{4x}-c_1,\\
&w_{5y}=\Phi_5w_{5x}+c_1,\ w_{5t}=9(3\Phi_5^3+7c_1\Phi_5)w_{5x}-c,\\
&w_{6y}=\Phi_6w_{6x}+c_1,\ w_{6t}=9(3\Phi_6^3+7c_1\Phi_6)w_{6x}-c_1,\\
\end{split}\end{eqnarray}
then the special combinations corresponding to 
\begin{equation}
w_7=w_1+w_2,\ w_8=w_3+\frac12w_4,\ w_9=\frac12(w_5+w_6)\label{SK7w789}
\end{equation}
are also solutions of the seventh-order PBKP equation \eqref{K7}. 

Substituting \eqref{SK7w789} with \eqref{w7pbkp} into \eqref{K7}, one can directly prove proposition 6. Therefore, in the seventh PBKP case, the principle of linear superposition \eqref{SK7w789} holds for the special types of decomposition solutions given by \eqref{w7pbkp}.

\subsection{Special linear superpositions of the ninth-order PBKP equation (\ref{K9})}

For the ninth-order PBKP equation \eqref{K9}, we have an analogous result for the linear superpositions of decomposition solutions.\\
\bf Proposition 7. \rm Let $w_1,\ w_2,\ w_3,\ w_4,\ w_5$ and $w_6$ be solutions of the ninth-order PBKP equation \eqref{K9} with the decompositions
\begin{eqnarray}\begin{split}
&w_{1y}=(\Phi_1+c) w_{1x},\ w_{1t}=(81\Phi_1^4+243c\Phi_1^3+243c^2\Phi_1^2
+90c^3 \Phi_1 +9c^4) w_{1x}-c,\\
&w_{2y}=(\Phi_2-c) w_{2x},\ w_{2t}=(81\Phi_2^4-243c\Phi_2^3+243c^2\Phi_2^2
-90c^3 \Phi_2 +9c^4) w_{2x}-c_1,\\
&w_{3y}=\Phi_3 w_{3x},\ w_{3t}=81\Phi_3^4 w_{3x}-c,\\
&w_{4y}=\Phi_4w_{4x},\ w_{4t}=81\Phi_4^4w_{4x}-c_1,
\\
&w_{5y}=\Phi_5w_{5x}+c_1,\ w_{5t}=\bigg(81\Phi_5^4+243c_1\Phi_5^2+\frac{405}2c_1^2\bigg)w_{5x}-c,
\\
&w_{6y}=\Phi_6w_{6x}+c_1,\ w_{6t}=\bigg(81\Phi_6^4+243c_1\Phi_6^2+\frac{405}2c_1^2\bigg)w_{6x}-c_1,
\\
\end{split}\end{eqnarray}
then $w_7=w_1+w_2,\ w_8=w_3+\frac12w_4$ and $w_9=\frac12(w_5+w_6)$ are still solutions of the ninth-order PBKP equation \eqref{K9}.

From the detailed examples of the fifth-, seventh-, and ninth-order PBKP equations, similar linear superposition conjecture can be found for the whole PBKP hierarchy \eqref{PBKPH}. 
\\
\bf Conjecture 3. \rm
For every equation of the PBKP hierarchy, there exist three possible special types of linear superposition solutions. 

It should be mentioned that though the solutions $w_i,\ i=1,\ \ldots,\ 6$ shown in the propositions 5, 6 and 7 are the potential KdV decomposition solutions, their special linear superposition solutions $w_7,\ w_8$ and $w_9$ are not the decomposition solutions. 

In the next section, we study the possible decompositions and linear superpositions for the dispersionless PBKP hierarchy.

\section{Decompositions and special linear superpositions of the dispersionless PBKP hierarchy}

\subsection{Mastersymmetries and dPBKP hierarchy}

The dPBKP equation possesses the form
\begin{equation}
w_{xt}=15(w_x w_{y}-w_x^3)_x+5w_{y2}= \partial_x\left(15w_xw_y-15w_x^3+5\partial_x^{-1}w_{yy}\right)\equiv \partial_x V.\label{dPBKP}
\end{equation}

Using the mastersymmetry approach similar to that described in conjecture 1 for the PBKP equation, one can make the following conjecture for the dPBKP equation \eqref{dPBKP}. \\
\bf Conjecture 4. \rm $y^n$ for arbitrary positive integers are mastersymmetries with $V_{[,]}^{n+1}y^n=0$.

If the conjecture is correct, then we can write the dPBKP hierarchy in the form
\begin{equation}
w_{t_{2n-1}}=V_{2n-1}=\frac{1}{3n!5^n}V_{[,]}^ny^n. \label{dPBKPH}
\end{equation}
We have check that the conjecture 4 is correct for $n=1,\ 2,\ \ldots,\ 6$. The corresponding equations of the dPBKP hierarchy \eqref{dPBKPH} possess the forms

\begin{eqnarray}
w_{t_{1}}&=&\frac{1}{15}V_{[,]}y=w_x, \label{V1}\\
w_{t_3}&=&\frac1{150}V_{[,]}^2y^2=3w_y, \label{V3}\\
w_{t_5}&=&\frac1{2250}V_{[,]}^3y^3=V_5=V, \label{V5}\\
w_{t_7}&=&\frac1{45000}V_{[,]}^4y^4=V_7\label{V7}\\
&=&21\partial_x^{-1}(w_xw_{y2}) +7\partial_x^{-2}w_{y3}-63w_x^2w_y +126\partial_x^{-1}(w_xw_{x2}w_y)
-63w_x^4+21w_x\partial_x^{-1}(w_{y2})+\frac{63}2w_y^2,\nonumber\\
w_{t_9}&=&\frac1{1125000}V_{[,]}^5y^5=V_9\label{V9}\\
&=&\partial_x^{-1}\bigg\{ \frac{27}2w_{x2} [15w_y^2-108w_x^2w_y+36\partial_x^{-1}(w_{x2}w_xw_y) +6\partial_x^{-1}(w_xw_{y2})+2\partial_x^{-2}w_{y3}]\nonumber\\
&&-81 w_{xy} (12w_x^3-5w_xw_y-\partial_x^{-1}w_{y2}) +135w_yw_{y2}+54w_x\partial_x^{-1}w_{y3} +27\partial_x^{-1}(w_xw_{y2})_y \nonumber\\
&&+162\partial_x^{-1}(w_{x2}w_xw_{y2}-w_xw_{xy}^2)+9\partial_x^{-2}w_{y4}\bigg\},\nonumber
\end{eqnarray}
\begin{eqnarray}
w_{t_{11}}&=&\frac1{33750000}V_{[,]}^6y^6=V_{11}\label{V11}\\
&=&\partial_x^{-2}\big\{w_{x3}[7128w_x^5+33\partial_x^{-3}w_{y4}-594\partial_x^{-2}(w_x w_{xy}^2) +99\partial_x^{-2}(w_{xy}w_{y2})+99\partial_x^{-2}(w_x w_{y3})\nonumber\\
&&+594\partial_x^{-1}(w_x w_{x2} w_{y2}) +198\partial_x^{-1}(w_{y}w_{y2})-99\partial_x^{-1}(w_{y3}w) +3564\partial_x^{-1}(w_x^2w_{x2}w_{y})-297\partial_x^{-1}(w_x^2w_{y2}) \nonumber\\
&& -9504w_{y}w_x^3-1485w_x^2\partial_x^{-1}w_{y2}+495w_{y}\partial_x^{-1}w_{y2} +99w\partial_x^{-1}w_{y3}] +w_{yx2}[2079/2w_{y}^2+99\partial_x^{-2}w_{y3}\nonumber\\
&&+1782\partial_x^{-1}(w_xw_{x2}w_{y}) +297\partial_x^{-1}(w_xw_{y2}) -3267w_x^4-2970w_x^2w_{y}+495 w_x\partial_x^{-1}w_{y2}]\nonumber\\
&&+w_{xy2}(1386 w_xw_{y}-1980 w_x^3 +165\partial_x^{-1}w_{y2})+(198w_x^2+231w_{y})w_{y3}+w_{x2}^2(35640 w_x^4-24948 w_x^2 w_{y}\nonumber\\
&& -2970 w_x \partial_x^{-1}w_{y2})+w_{x2}[w_{xy}(990\partial_x^{-1}w_{y2}-22572w_x^3-4158w_xw_{y}) +w_{y2}(2079w_{y}-5346w_x^2)\nonumber\\
&&+297w_x\partial_x^{-1}w_{y3}+297\partial_x^{-1}(w_{xy} w_{y2} )+297\partial_x^{-1}(w_xw_{y3}) -1782\partial_x^{-1}(w_xw_{xy}^2)+1782\partial_x^{-1}(w_xw_{x2}w_{y2}) \nonumber\\
&& +99\partial_x^{-2}w_{y4}]+w_{xy}^2(-5346 w_x^2+2079 w_{y})+w_{xy}(1782w_xw_{y2}+297\partial_x^{-1}w_{y3}) +99w_x\partial_x^{-1}w_{y4}\nonumber\\
&&+33\partial_x^{-1}w_xw_{y4} +11\partial_x^{-2}w_{y5}
+\frac{495}2 w_{y2}^2-198\partial_x^{-1}w_{xy}^3+594\partial_x^{-1}(w_{y2}w_{x2}w_{xy}) +594\partial_x^{-1}(w_{y2}w_xw_{yx2})\nonumber\\
&&+198\partial_x^{-1}(w_xw_{x2}w_{y3}) +66\partial_x^{-1}(w_{xy}w_{y3})\big\}.\nonumber
\end{eqnarray}

\subsection{Decompositions of dPBKP hierarchy}

\bf Proposition 8. \rm If $w_1,\ w_2,\ w_3$ and $w_4$ are solutions of the following dispersionless KdV systems,
\begin{eqnarray}
&&\left\{\begin{array}{l}
\displaystyle{w_{1y}=a_2w_{1x}^2+a_1 w_{1x}+c_1},\\
\\
\displaystyle{w_{1t_5}=\frac53(a_2+3)(4a_2-3)w_{1x}^3+5a_1(2a_2+3)w_{1x}^2+5(a_1^2+3c_1)w_{1x}},
\end{array}\right. \label{dC51}\\
&&\left\{\begin{array}{l}
\displaystyle{w_{2y}=a_2w_{2x}^2+a_1 w_{2x}+c_1},\\
\\
\displaystyle{w_{2t_7}=\frac72 (a_2-1) (4 a_2+3) (a_2+6)w_{2x}^4+7a_1(4a_2^2+19 a_2-3)w_{2x}^3 +b w_{2x}^2+7a_1(a_1^2+9c_1)w_{2x}},
\end{array}\right. \label{dC52}\\
&&\left\{\begin{array}{l}
\displaystyle{w_{3y}=a_2w_{3x}^2+a_1 w_{3x}+c_1},\\
\\
\displaystyle{w_{3t_9}=\frac9{10} a_2 (a_2+9) (8 a_2-9) (4 a_2+9)w_{3x}^5 +c_4w_{3x}^4 +c_3 w_{3x}^3 +c_2w_{3x}^2+\frac92 c_1w_{3x}},
\end{array}\right. \label{dC53}
\\
&&\left\{\begin{array}{l}
\displaystyle{w_{4y}=a_2w_{4x}^2+a_1 w_{4x}+c_1},\\
\\
\displaystyle{w_{4t_{11}}=\frac{11}{30}(d_6w_{4x}^6+3d_5w_{4x}^5 +15 d_4w_{4x}^4 +10 d_3 w_{4x}^3 +15d_2w_{4x}^2+15d_1w_{4x})},
\end{array}\right. \label{dC54}
\end{eqnarray}
with $b=21(a_1^2a_2+3c_1a_2+3a_1^2)$, $c_4=9a_1(2a_2+3)(4a_2^2+30a_2-21)$, $c_3=9(8a_1^2a_2^2+20c_1a_2^2+58a_1^2a_2+45c_1a_2+24a_1^2-45c_1)$, $c_2=9a_1(4a_1^2a_2+39c_1a_2+18a_1^2+45c_1)$, $c_1=2a_1^4+54c_1a_1^2+45c_1^2$, $d_6=(a_2+12) (5 a_2-6) (4 a_2+15) (8 a_2-3) (a_2+1)$, $d_5=a_1 (160 a_2^4+2236 a_2^3+4551 a_2^2-3357 a_2-2160)$, $d_4=(40 a_1^2 a_2^3+84 c_1 a_2^3+490 a_1^2 a_2^2 +483 c_1 a_2^2+780 a_1^2 a_2-189 c_1 a_2-270 a_1^2-378 c_1)$, $d_3=a_1 (40 a_1^2 a_2^2+252 c_1 a_2^2+390 a_1^2 a_2+1197 c_1 a_2+360 a_1^2-189 c_1)$, $d_2=10 a_1^4 a_2+144 c_1 a_1^2 a_2+60 a_1^4+189 c_1^2 a_2+378 c_1 a_1^2)$ and $d_1=a_1 (2 a_1^4+66 c_1 a_1^2+189 c_1^2)$, then $w_1,\ w_2,\ w_3$ and $w_4$ are solutions of the fifth-, seventh-, ninth- and eleventh-order dPBKP equations \eqref{V5}-\eqref{V11}, respectively.

In fact, for the dPBKP hierarchy there are more general decomposition solutions than those shown in proposition 8. Here is a general decomposition conjecture:\\
\bf Conjecture 5. \rm The dPBKP hierarchy \eqref{dPBKPH} possesses a $\{y,\ t_{2n-1}\}$ decomposed solution $w_n$ for the $n$th-order equation
\begin{eqnarray}
\left\{\begin{array}{l}
\displaystyle{w_{ny}=F(w_{nx})},\\
\\
\displaystyle{w_{nt_{2n-1}}=G(w_{nx})}
\end{array}\right. \label{dCn}
\end{eqnarray}
with $F(w_{nx})$ being an arbitrary function of $w_{nx}$ and $G(w_{nx})$ being a function determined by $F(w_{nx})$. Especially, if
\begin{equation}\label{Fn}
F(w_{nx})=\sum_{k=0}^ma_kw_{nx}^k,
\end{equation}
then
\begin{equation}\label{Gn}
G(w_{nx})=\sum_{k=0}^{(m-1)(n-1)+1}b_kw_{nx}^k
\end{equation}
with $b_k$ being determined by $a_k$.

For the fifth-order dPBKP equation \eqref{V5}, $n=3$, it is easy to find that
$$G(x_3)=-15x_3^3+15x_3F+5\partial_{x_3}^{-1}F_{x_3}^2, \ F=F(x_3),\ x_3=w_{3x}.$$
For the seventh-order dPBKP equation \eqref{V7}, $n=4$, we have
$$G(x_4)=-\frac{21}2\big[6x_4^4-3F^2+\partial_{x_4}^{-1}(6x_4^2F_{x_4}-4x_4F_{x_4}^2)\big], \ F=F(x_4),\ x_4=w_{4x}.$$

Proposition 8 is just for the special cases of conjecture 5 related to $m=2$ of \eqref{Fn} and \eqref{Gn}.

\subsection{Linear superpositions of dPBKP hierarchy}

Similar to the PBKP hierarchy \eqref{PBKPH}, for some special decomposition solutions, the dPBKP hierarchy also admits solutions in terms of their sum.
For the fifth-, seventh-, ninth- and eleventh-order dPBKP equations \eqref{V5}-\eqref{V11}, some special linear superpositions are summarized in the following propositions 9, 10, 11 and 12. \\
\bf Proposition 9. \rm Assume $w_1,\ w_2,\ w_3,\ w_4,\ w_5$ and $w_6$ are solutions of the fifth-order dPBKP equation \eqref{V5} with the decomposition conditions
\begin{eqnarray}
&&w_{1y}=3w_{1x}^2+c w_{1x},\
w_{1t_5}=90w_{1x}^3+45c w_{1x}^2+5c^2 w_{1x},\\
&&w_{2y}=3w_{2x}^2-c w_{2x},\
w_{2t_5}=90w_{2x}^3-45c w_{2x}^2+5c^2 w_{2x},\\
&&w_{3y}=3w_{3x}^2,\ w_{3t_5}=90w_{3x}^3,\\
&&w_{4y}=\frac32w_{4x}^2,\ w_{4t_5}=\frac{45}2w_{4x}^3,\\
&&w_{5y}=\frac32w_{5x}^2+c_1,\ w_{5t_5}=\frac{45}2w_{5x}^3+15c_1w_{5x},\\
&&w_{6y}=\frac32w_{6x}^2+c_1,\ w_{6t_5}=\frac{45}2w_{6x}^3+15c_1w_{6x},
\end{eqnarray}
then $w_7$, $w_8$ and $w_9$ with the linear superposition properties
\begin{equation}
w_7=w_1+w_2,\ w_8=w_3+\frac12w_4,\ w_9=\frac12(w_5+w_6)
\end{equation}
are also solutions of the fifth-order dPBKP equation \eqref{V5}.
\\
\bf Proposition 10. \rm If the solutions $w_i,\ i=1,\ 2,\ \ldots,\ 12$, of the seventh-order dPBKP equation \eqref{V7} are satisfied by the decomposition properties 

\begin{eqnarray}
&&w_{1y}=3w_{1x}^2+c w_{1x},\
w_{1t_7}=7\big(135w_{1x}^4+90c w_{1x}^3+18c^2 w_{1x}^2+c^3 w_{1x}\big),\\
&&w_{2y}=3w_{2x}^2-c w_{2x},\
w_{2t_7}=7\big(135w_{2x}^4-90c w_{2x}^3+18c^2 w_{2x}^2-c^3 w_{2x}\big),\\
&&w_{3y}=3w_{3x}^2,\ w_{3t_7}=945w_{3x}^4,\\
&&w_{4y}=\frac32w_{4x}^2,\ w_{4t_7}=\frac{945}8w_{4x}^4,\\
&&w_{5y}=\frac32w_{5x}^2+c_1,\ w_{5t_7}=\frac{945}8w_{5x}^4+\frac{189}2c_1w_{5x}^2,\\
&&w_{6y}=\frac32w_{6x}^2+c_1,\ w_{6t_7}=\frac{945}8w_{6x}^4+\frac{189}2c_1w_{6x}^2,\\
&&w_{7y}=3w_{7x}^2,\ w_{7t_7}={945}w_{7x}^4,\\
&&w_{8y}=-\frac23w_{8x}^2,\ w_{8t_7}=-\frac{280}{27}w_{8x}^4,\\
&&w_{9y}=-\frac23w_{9x}^2,\ w_{9t_7}=-\frac{280}{27}w_{9x}^4,\\
&&w_{10y}=-\frac23w_{10x}^2,\ w_{10t_7}=-\frac{280}{27}w_{10x}^4,\\
&&w_{11y}=-\frac23w_{11x}^2,\ w_{11t_7}=-\frac{280}{27}w_{11x}^4,\\
&&w_{12y}=\frac32w_{12x}^2,\ w_{12t_7}=\frac{945}{8}w_{12x}^4,
\end{eqnarray}
then $w_{i},\ i=13,\ \ldots,\ 18$, with the linear superposition behaviors
\begin{eqnarray}
&&w_{13}=w_1+w_2,\ w_{14}=w_3+\frac12w_4,\ w_{15}=\frac12(w_5+w_6),\nonumber\\
&& w_{16}=w_7-\frac29w_8,\
w_{17}=-\frac29(w_9+w_{10}),\ w_{18}=-\frac29w_{11}+\frac12w_{12}
\end{eqnarray}
are also solutions of the seventh-order dPBKP equation \eqref{V7}.
\\
\bf Proposition 11. \rm The solutions $w_i,\ i=1,\ 2,\ \ldots,\ 12$,  with the decomposition relations
\begin{eqnarray}
&&w_{1y}=3w_{1x}^2+c w_{1x},\
w_{1t_9}=9w_{1x}(6w_{1x}+c)(3w_{1x}+c)(63w_{1x}^2+21c w_{1x}+c^2),\\
&&w_{2y}=3w_{2x}^2-c w_{2x},\
w_{2t_9}=9w_{2x}(6w_{2x}-c)(3w_{2x}-c)(63w_{2x}^2-21c w_{2x}+c^2),\\
&&w_{3y}=3w_{3x}^2,\ w_{3t_9}=10206w_{3x}^5,\\
&&w_{4y}=\frac32w_{4x}^2,\ w_{4t_9}=\frac{5103}8w_{4x}^5,\\
&&w_{5y}=\frac32w_{5x}^2+c_1,\ w_{5t_9}=\frac{5103}8w_{5x}^5+\frac{1215}2c_1w_{5x}^3 +\frac{405}2c_1^2w_{5x},\\
&&w_{6y}=\frac32w_{6x}^2+c_1,\ w_{6t_9}=\frac{5103}8w_{6x}^5+\frac{1215}2c_1w_{6x}^3 +\frac{405}2c_1^2w_{6x},\\
&&w_{7y}=3w_{7x}^2,\ w_{7t_9}={10206}w_{7x}^5,\\
&&w_{8y}=-\frac32w_{8x}^2,\ w_{8t_9}=\frac{5103}{8}w_{8x}^5,\\
&&w_{9y}=-\frac32w_{9x}^2+c_1,\ w_{9t_9}=\frac{5103}8w_{9x}^5-\frac{1215}2c_1w_{9x}^3+\frac{405}2c_1^2w_{9x},\\
&&w_{10y}=-\frac32w_{10x}^2+c_1,\ w_{10t_9}=\frac{5103}8w_{10x}^5-\frac{1215}2c_1w_{10x}^3+\frac{405}2c_1^2w_{10x},\\
&&w_{11y}=-\frac32w_{11x}^2+c_1,\ w_{11t_9}=\frac{5103}8w_{11x}^5-\frac{1215}2c_1w_{11x}^3+\frac{405}2c_1^2w_{11x},\\
&&w_{12y}=\frac32w_{12x}^2+c_1,\ w_{12t_9}=\frac{5103}8w_{12x}^5+\frac{1215}2c_1w_{12x}^3+\frac{405}2c_1^2w_{12x}
\end{eqnarray}of the ninth-order dPBKP equation \eqref{V9} give rise to 
the linear superposition solutions 
\begin{eqnarray}
&&w_{13}=w_1+w_2,\ w_{14}=w_3+\frac12w_4,\ w_{15}=\frac12(w_5+w_6),\nonumber\\
&& w_{16}=w_7-\frac12w_8,\
w_{17}=-\frac12(w_9+w_{10}),\ w_{18}=-\frac12w_{11}+\frac12w_{12}
\end{eqnarray}of the same equation.
\\
\bf  Proposition 12. \rm For any solution $w_i,\ i=1,\ 2,\ \ldots,\ 13,$ satisfying decomposition conditions
\begin{eqnarray}
&&w_{1y}=3w_{1x}^2+c w_{1x},\
w_{1t_{11}}=11w_{1x}(c+3 w_{1x}) (3402 w_{1x}^4+2268 c w_{1x}^3+504c^2 w_{1x}^2+42c^3 w_{1x}+c^4),\\
&&w_{2y}=3w_{2x}^2-c w_{2x},\
w_{2t_{11}}=11w_{2x}(3 w_{2x}-c) (3402 w_{2x}^4-2268 c w_{2x}^3+504c^2 w_{2x}^2-42c^3 w_{2x}+c^4),\\
&&w_{3y}=3w_{3x}^2,\ w_{3t_{11}}=112266w_{3x}^6,\\
&&w_{4y}=\frac32w_{4x}^2+c_1,\ w_{4t_{11}}=\frac{6237}{16}w_{4x}^2(9w_{4x}^4+10 c_1w_{4x}^2+4c_1^2),\\
&&w_{5y}=\frac32w_{5x}^2+c_1,\ w_{5t_{11}}=\frac{6237}{16}w_{5x}^2(9w_{5x}^4+10 c_1w_{5x}^2+4c_1^2),\\
&&w_{6y}=-\frac32w_{6x}^2+c_1,\ w_{6t_{11}}=-\frac{6237}{16}w_{6x}^2(9w_{6x}^4-10 c_1w_{6x}^2+4c_1^2),\\
&&w_{7y}=-\frac32w_{7x}^2+c_1,\ w_{7t_{11}}=-\frac{6237}{16}w_{7x}^2(9w_{7x}^4-10 c_1w_{7x}^2+4c_1^2),\\
&&w_{8y}=\frac32w_{8x}^2,\ w_{8t_{11}}=\frac{56113}{16}w_{8x}^6,\\
&&w_{9y}=-\frac32w_{9x}^2,\ w_{9t_{11}}=-\frac{56113}{16}w_{9x}^6,\\
&&w_{10y}=6 c_+ w_{10x}^2,\ w_{10t_{11}}=\frac{4032(503275411c_{+} -30935290)}{519921875}w_{10x}^6,\ c_{\pm}=\frac{-109 \pm \sqrt{44881}}{1605},\\
&&w_{11y}=6 c_+ w_{11x}^2,\ w_{11t_{11}}=\frac{4032(503275411c_{+} -30935290)}{519921875}w_{11x}^6,\\
&&w_{12y}=6 c_{-} w_{12x}^2,\ w_{12t_{11}}=\frac{4032(503275411c_{-} -30935290)}{519921875}w_{12x}^6,\\
&&w_{13y}=6 c_{-} w_{13x}^2,\ w_{13t_{11}}=\frac{4032(503275411c_{-} -30935290)}{519921875}w_{13x}^6,
\end{eqnarray}
there exist the linear superposition solutions 
\begin{eqnarray}
&&w_{14}=w_1+w_2,\ w_{15}=\frac12(w_4+w_5),\ w_{16}=-\frac12(w_6+w_7),\ w_{17}=\frac12(w_4-w_6),\nonumber\\
&& w_{18}=w_3+\frac12w_8,\
w_{19}=w_3-\frac12w_9,\ w_{20}=w_3+2c_{+}w_{10},\ w_{21}=w_3+2c_{-}w_{12},\nonumber\\
&&w_{22}=\frac12 w_8+2c_{+}w_{10}, \ w_{23}=\frac12 w_8+2c_{-}w_{12}, \
w_{24}=2c_{+}w_{10}-\frac12 w_9, \ w_{25}=2c_{-}w_{12}-\frac12 w_9, \nonumber\\
&&w_{26}=2c_{+}w_{10}+2c_{+}w_{11}, \ w_{27}=2c_{+} w_{10}+2c_{-}w_{12}, \
w_{28}=2c_{-} w_{12}+2c_{-}w_{13}
\end{eqnarray}
of the eleventh-order dPBKP equation \eqref{V11}.
\\
\bf Conjecture 6. \rm As the orders of the dPBKP equations increase, the possible types of linear superposition solutions increase.

Same as in the PBKP hierarchy cases, the special linear superposition solutions of the dPBKP equations are not decomposition solutions.


\section{Conclusions and discussions}


We analyze the special decompositions and some linear superpositions of the BKP hierarchy and the dispersionless BKP hierarchy from their potential forms which are constructed in terms of the mastersymmetry approach.
For the PBKP hierarchy, we conjecture that every order PBKP equation admits six different types of decomposition solutions. Particularly, we prove this conjecture for three lower-order PBKP equations. 
These decompositions provide simple and surprising relationships between several classic integrable systems and the PBKP hierarchy.
The interrelations between different integrable systems enable us to discover unexpected connections of various mathematical and physical problems, ultimately solve them. 
Another interesting result about these special decomposition solutions is their suitable linear superpositions can also yield new solutions of the PBKP hierarchy, although the linear superposition theorem does not apply to nonlinear systems in general. 
This is equivalent to say that a possible linear combination of two special types of decomposition solutions holds again a solution to the same equation.
The obtained linear superposition formulas allow us to construct many new exact solutions including $m$+$n$ solitons, $n$ solitons with periodic cnoidal background waves, $n$ solitons combined with soliton-cnoidal wave interaction solutions, and combination solutions of two different modified Schwarzian KdV waves, etc. of the PBKP hierarchy.
We mention that only two of the decomposition solutions can be linearly combined to construct three new possible solutions, although there are six possible decompositions for all equations in the PBKP hierarchy.
The correctness of this has been verified through the fifth-, seventh-, and ninth-order PBKP equations.

Furthermore, we propose conjectures about the existence of decompositions and linear superposition solutions for each equation in the dPBKP hierarchy and, particularly, verify these conjectures for the fifth-, seventh-, ninth-, and eleventh-order dPBKP equations. 
Different from that which is common to all the members of the PBKP hierarchy containing the same possible linear superpositions, the possible ways of linear superposition increase rapidly as the order of the dPBKP equation increases.


We also emphasize that 
the significance of our findings about linear superposition properties is not restricted to the PBKP hierarchy or the dispersionless PBKP hierarchy. Our findings confirm the existence of some possible special linear superposition solutions in nonlinear systems and add the richness of exact solutions. Particularly, the existence of such linear superposition solutions in nonlinear systems provides us a totally new insight into the physical nature.


Naturally, generalizing the linear superposition principle, we will further consider whether possible linear superpositions of three or more special types of decomposition solutions exist in the PBKP hierarchy or dPBKP hierarchy. 
And is it possible to find linear superposition properties in other wonderful properties such as Hirota's bilinear representations, Schwarzian forms, etc. of the integrable systems?
There is much remaining to be explored about how linear superposition principle can challenge and enrich our understanding of nonlinear systems. 
These, and other issues concerning linear superpositions, merit investigation.

\section*{Declaration of Competing Interest}
Authors declare that they have no conflict of interest.

\section*{Acknowledgement}
The work was sponsored by the National Natural Science Foundations of China (Nos. 11975131, 11435005), K. C. Wong Magna Fund in Ningbo University, the Natural Science Foundation of Zhejiang Province No. LQ20A010009 and the General Scientific Research of Zhejiang Province No. Y201941009. The authors would like to thank Professors X. B. Hu and Q. P. Liu for their valuable discussions.

\bibliography{Myrefs}

\end{CJK*}
\end{document}